%% file: ANA-SUSY-2019-12-PAPER.tex
\documentclass[cernpreprint, texlive=2023, UKenglish, texmf, orcidlogo]{atlasdoc}
 
\usepackage[subcaption]{atlaspackage}
\usepackage{atlasbiblatex}
 
\usepackage{float}
\usepackage{multirow}
 
\usepackage[hepprocess]{atlasphysics}
 
\graphicspath{{logos/}{figures/}}
 
\usepackage{ANA-SUSY-2019-12-PAPER-defs}

\AtlasTitle{Constraints on spin-0 dark matter mediators and invisible Higgs decays using ATLAS 13~TeV $pp$ collision data with two top quarks and missing transverse momentum in the final state}

\PreprintIdNumber{CERN-EP-2022-133}
 
\AtlasAbstract{
This paper presents a statistical combination of searches targeting final states with two top quarks and invisible particles, characterised by the presence
of zero, one or two leptons, at least one jet originating from a $b$-quark and missing transverse momentum.
The analyses are searches for phenomena beyond the Standard Model consistent with the direct production of dark matter in $pp$ collisions at the LHC, using 139~fb$^{-\text{1}}$ of data collected with the ATLAS detector at a centre-of-mass energy of 13~TeV.
The results are interpreted in terms of simplified dark matter models with a spin-0 scalar or pseudoscalar mediator particle. In addition,
the results are interpreted in terms of upper limits on the Higgs boson invisible branching ratio, where the Higgs boson is produced
according to the Standard Model in association with a pair of top quarks.
For scalar (pseudoscalar) dark matter models, with all couplings set to unity, the statistical combination extends the mass range
excluded by the best of the individual channels by 50~(25)~GeV, excluding mediator masses up to 370~GeV.
In addition, the statistical combination improves the expected coupling exclusion reach by 14\% (24\%), assuming a scalar (pseudoscalar) mediator mass of 10~GeV.
An upper limit on the Higgs boson invisible branching ratio of 0.38 ($\text{0.30}^{+\text{0.13}}_{-\text{0.09}}$) is observed (expected) at 95\% confidence level.
}

\author{The ATLAS Collaboration}
 
\AtlasRefCode{SUSY-2019-12}

\AtlasJournal{EPJC}
\AtlasJournalRef{Eur. Phys. J. C 83 (2023) 503}
\AtlasDOI{10.1140/epjc/s10052-023-11477-z}

\hypersetup{pdftitle={ATLAS document},pdfauthor={The ATLAS Collaboration}}
 
\begin{document}
 
\maketitle
 
\tableofcontents

\section{Introduction}

The existence of a non-luminous component of matter in the universe, dark matter (DM), is
supported by compelling astrophysical evidence~\cite{Zwicky:1933gu, DMreview}.
The abundance of dark matter has been precisely determined
from global
fits of cosmological parameters to a variety of observations~\cite{Komatsu:2010fb,Planck:2015fie}.
Nevertheless, the nature of DM remains largely unknown.
 
In this paper, models where the DM candidate is a
weakly interacting massive particle (WIMP)~\cite{Steigman:1984ac} are considered.
In proton--proton ($pp$) collisions at the Large Hadron Collider (LHC), pair-produced WIMP DM
does not interact with the detector
and it can be detected only if produced in association with Standard Model (SM) particles,
leading to signatures with missing transverse momentum.
A wide range of experimental searches
are focused on WIMP candidates at the LHC~
\cite{EXOT-2018-06,CMS-EXO-19-003,CMS-EXO-19-003,CMS-EXO-19-003,CMS-EXO-19-003,SUSY-2018-34,EXOT-2018-63,SUSY-2016-18,SUSY-2016-18,ATLAS:2021shl,EXOT-2017-32}.
All recent searches are based on simplified benchmark models
documented in the LPCC Dark Matter Working Group whitepapers~\cite{Boveia:2016mrp,Albert:2017onk,LHCDarkMatterWorkingGroup:2018ufk,Abercrombie:2015wmb}.
Benchmark models are chosen to have a minimal number of additional parameters relative to the SM.
This paper focuses on simplified models characterised by the introduction of a spin-0 particle mediator~\cite{Blanke:2020bsf,Westhoff:2019pzy,LHCDarkMatterWorkingGroup:2018ufk,Haisch:2018bby,Haisch:2016gry,Haisch:2015ioa,Abercrombie:2015wmb,Abdallah:2015ter,Buckley:2014fba}. In this case, fermionic DM particle pairs
are produced through the exchange of a colour-neutral scalar or
pseudoscalar mediator (denoted by $\phi$ or $a$, respectively).
In the Minimal Flavour
Violation~\cite{DAmbrosio:2002vsn} assumption,
the interaction between any new neutral spin-0
state and SM quarks is proportional to the fermion masses via
Yukawa-type couplings and it is also treated as a free parameter of the model by means of a multiplicative factor $g_q$.
Following Ref.~\cite{Buckley:2014fba}, couplings to leptons and $W$/$Z$ bosons, as well as explicit $\phi$--$H$ or $a$--$H$ couplings of dimension four to the SM Higgs boson,
are set to zero. The coupling of the mediator to the dark sector, $g_{\chi}$, is not assumed to be proportional to the mass of
the DM candidates and is treated simply as a free parameter.
 
\begin{figure}[ht!b]
\begin{center}
\begin{subfigure}[t]{0.22\textwidth}
\centering
\includegraphics[width=0.99\textwidth]{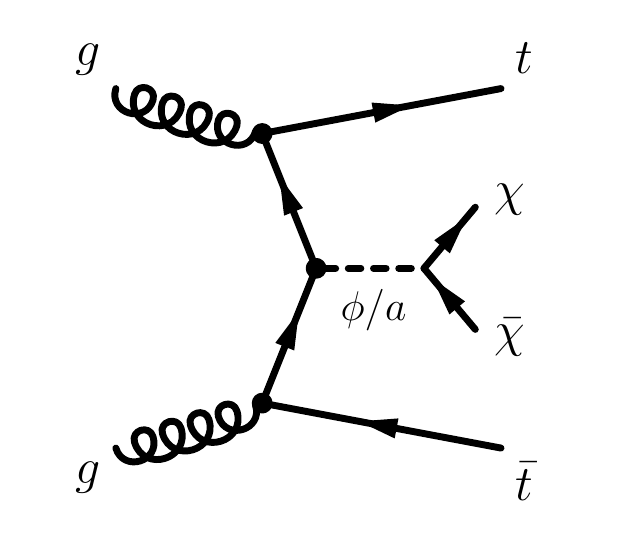}
\caption{}
\end{subfigure}
\begin{subfigure}[t]{0.22\textwidth}
\centering
\includegraphics[width=0.99\textwidth]{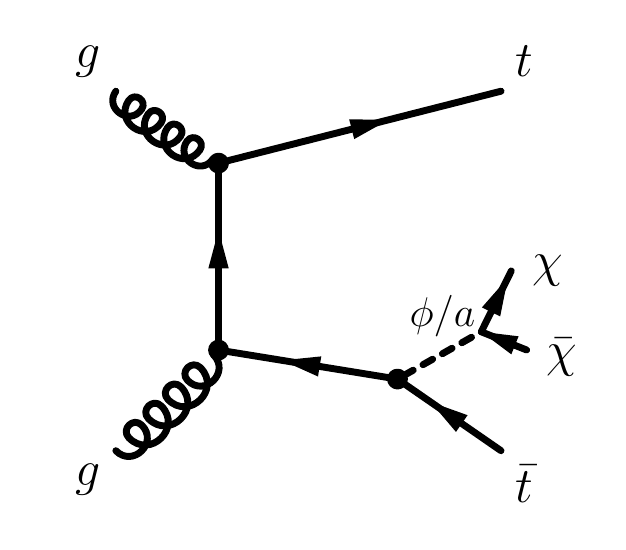}
\caption{}
\end{subfigure}
\begin{subfigure}[t]{0.22\textwidth}
\centering
\includegraphics[width=0.99\textwidth]{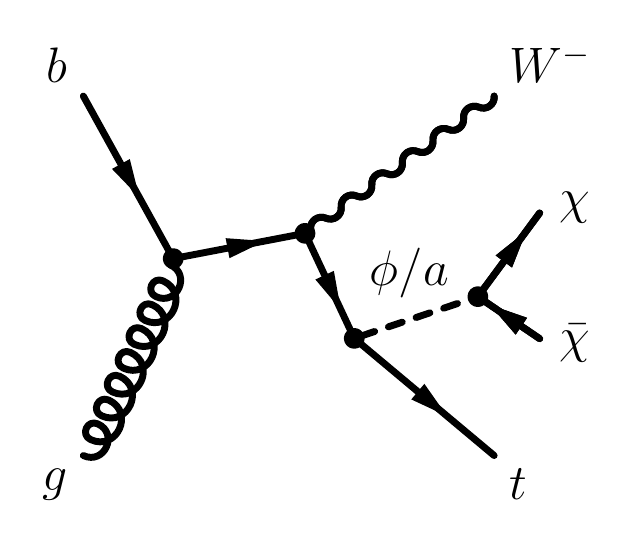}
\caption{}
\end{subfigure}
\begin{subfigure}[t]{0.22\textwidth}
\centering
\includegraphics[width=0.99\textwidth]{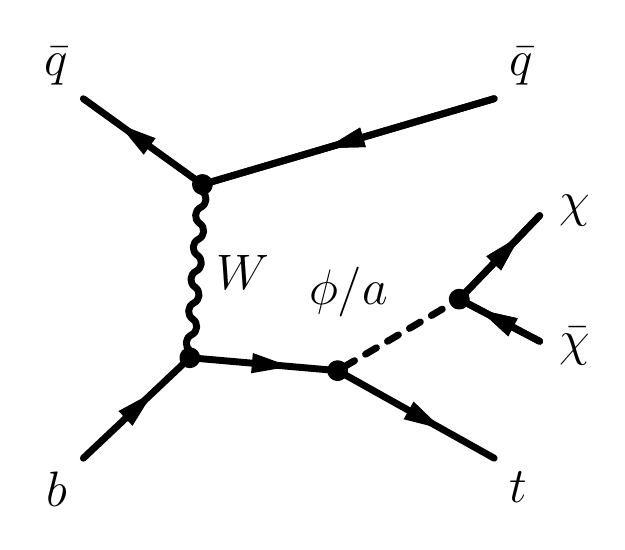}
\caption{}
\end{subfigure}
\caption{Representative diagrams for spin-0 mediator associated production with (a) and (b) a top quark pair (DM+$\ttbar$),
(c) a single top quark and a $W$ boson (DM+$tW$) or (d) a single top quark and one (or more) jet(s) (DM+$tj$).
}
\label{fig:diagrams}
\end{center}
\end{figure}

The dominant production modes for such colour-neutral mediators are
loop-induced gluon-fusion and associated production of the mediator with a top quark pair.
Figures~\ref{fig:diagrams}(a) and~\ref{fig:diagrams}(b) show the two dominant production diagrams for the associated production of the mediator with a top quark pair.
As discussed in Refs.~\cite{Haisch:2016gry,Backovic:2015soa}, the diagram in Figure~\ref{fig:diagrams}(b) can be interpreted as the radiation of the mediator from a top quark. This process is enhanced for small energies and leads to a production cross section for scalar mediators about one order of magnitude larger than the one for pseudoscalar mediators for masses below the top quark threshold.
The associated production of spin-0 mediators with a single top quark
also has a sizeable, albeit non-dominant, cross section~\cite{Pinna:2017tay,Pani:2017qyd,Haisch:2018bby}, especially for higher-mass mediators.
The primary signal of interest in this paper is the associated production of a mediator particle with a pair of top quarks (DM+$\ttbar$),
although sensitivity is also retained in the case of single top quark production (collectively referred to as DM+$t$).
The relevant processes for DM+$t$ are shown in Figures~\ref{fig:diagrams}(c) and~\ref{fig:diagrams}(d).
The relative contribution of the two processes depends on the parameter space that is considered~\cite{Pinna:2017tay}.
 
This paper presents a statistical combination of three searches targeting events with two top quarks and invisible particles,
considering
either zero- (tt0L~\cite{SUSY-2018-12}), one- (tt1L~\cite{SUSY-2018-07}) or two-lepton (tt2L~\cite{SUSY-2018-08}) final states,
using 139~fb$^{-1}$ of $pp$ collisions data recorded by the ATLAS detector~\cite{PERF-2007-01, ATLAS-TDR-2010-19} at a centre-of-mass energy of $\sqrt{s}
= 13~\TeV$.
The tt0L analysis is extended and improved relative to that in Ref.~\cite{SUSY-2018-12},
benefitting from the Run-2 improvements in the trigger selection of jets containing
$b$-hadrons ($b$-jets) in ATLAS~\cite{ATLAS:2021piz} that increase the sensitivity in models with low-mass mediators.
The three analyses select independent (orthogonal) datasets which can be statistically combined to boost the sensitivity
because they seek signal events in similar parts of the parameter space.
Searches specifically targeting the associated
production of a single top quark and invisible particles \cite{ATLAS:2020yzc} are not orthogonal to
the searches presented here. They are therefore not included and their combination is left as a future development.
 
The data are found to be in agreement with the SM background prediction in all selections considered in
this paper. The statistical combination is used
to set 95\% confidence level (CL) constraints on the simplified DM
models.
The specific case where the mediator corresponds to the SM $125~\GeV$ Higgs boson~\cite{HIGG-2012-27}
is also considered when interpreting the results presented in this paper. It is referred to as the \hinv\ interpretation in the following. In
the SM, the invisible Higgs boson branching ratio, \brinv,  is 0.12\% from $H \rightarrow ZZ \rightarrow  4\nu$ decays~\cite{deFlorian:2016spz},
and higher branching ratios to invisible particles are predicted by Higgs--dark-matter portal models~\cite{Shrock:1982kd,Choudhury:1993hv,ArkaniHamed:1998vp,Eboli:2000ze,Godbole:2003it,Antoniadis:2004se,Davoudiasl:2004aj,Patt:2006fw,Kanemura:2010sh,Djouadi:2011aa,Ghosh:2012ep,Belanger:2013kya,Curtin:2013fra,Zaazoua:2021xls}.
Results on the invisible branching ratio obtained from the statistical combination of previous ATLAS searches conducted with the Run-1 and partial Run-2 dataset reported an observed (expected) limit of $0.26$ ($0.17$)~\cite{HIGG-2018-54}, while CMS reported $0.19$ ($0.15$)~\cite{CMS-HIG-17-023} at $95\%$ CL.
Recent updates by the ATLAS and CMS Collaborations in the vector-boson-fusion channel report improved observed (expected) upper limits of $0.145~(0.103)$ \cite{ATLAS:2022yvh} and $0.18~(0.10)$~\cite{PhysRevD.105.092007}, respectively, using the full Run-2 dataset.
 
The paper is structured as follows. The experimental dataset and the simulated event samples are presented in Section~\ref{sec:samples}, the new and previously published searches are introduced in Section~\ref{sec:analysis}, and their statistical combination, including the treatment of correlated systematic uncertainties is discussed in Section~\ref{sec:comb}. Results in terms of exclusion limits are presented in Section~\ref{sec:results}, followed by the conclusions in Section~\ref{sec:conclusion}. Finally, additional details of the extended 0-lepton channel analysis are discussed in Appendix~\ref{sec:lowmetapp}.


\section{Data and simulated event samples}
\label{sec:samples}

 
The dataset used in the analyses described in this paper consists of
$pp$ collision data recorded by the ATLAS detector~\cite{PERF-2007-01,ATLAS-TDR-2010-19} at a centre-of-mass energy of $\sqrt{s}
= 13~\TeV$ with stable beam conditions.
The ATLAS detector is a multipurpose particle detector with a forward--backward symmetric cylindrical geometry and nearly full coverage in solid angle.\footnote{
ATLAS uses a right-handed coordinate system with its origin at the nominal interaction point (IP) in the centre of the detector and the $z$-axis
along the beam pipe. The $x$-axis points from the IP to the centre of the LHC ring, and the $y$-axis points upward.
Cylindrical coordinates $(r,\phi)$ are used in the transverse plane, $\phi$ being the azimuthal angle around the $z$-axis.
The pseudorapidity is defined in terms of the polar angle $\theta$ as $\eta=-\ln\tan(\theta/2)$.
The angular distance between two objects in $\eta$--$\phi$ space is defined by $\Delta R \equiv \sqrt{(\Delta\eta)^2+(\Delta\phi)^2}$.}
It consists of an inner tracking detector surrounded by a thin superconducting solenoid providing a 2~T axial magnetic field, electromagnetic and hadron calorimeters, and a muon spectrometer.
An extensive software suite~\cite{ATL-SOFT-PUB-2021-001} is used in data simulation, in the reconstruction and analysis of real and simulated data, in detector operations, and in the trigger and data acquisition systems of the experiment.
 
All collision events considered in this paper are required to have at
least one reconstructed interaction vertex with a minimum of two
associated tracks, each having $\pt>500$~\MeV.
In events with multiple vertices, the one with the highest sum of
squared transverse momenta of associated tracks is
chosen as the primary vertex~\cite{ATL-PHYS-PUB-2015-026}.
Electrons~\cite{EGAM-2018-01}, muons~\cite{PERF-2015-10}, $\tau$-leptons~\cite{PERF-2013-06} and jets are reconstructed by combining the
signals from the different components of the ATLAS
detector.
Small-radius jets are reconstructed from energy deposits in the calorimeters using
the anti-$k_t$ jet
algorithm~\cite{paper:antikt,Cacciari:2011ma} with a radius parameter
of $R = 0.4$. Reclustered large-radius jets \cite{Nachman:2014kla,ATLAS-CONF-2017-062} are reconstructed
with the same algorithm, using a radius parameter of $R=1.2$ and with small-radius jets as input, unless otherwise specified.
Multivariate algorithms  are used to
identify
small-$R$ jets with $\pt >20\;\GeV$ containing $b$-hadrons
($b$-jets)~\cite{PERF-2012-04,ATL-PHYS-PUB-2016-012}.
This is referred to as $b$-tagging.
The missing transverse momentum $\vec{p}_{\text{T}}^{\text{miss}}$ (with
magnitude $\met$) is calculated from the negative vector sum of the transverse
momenta  of electrons, muons
and jet candidates, and an additional `soft term'~\cite{PERF-2016-07}
which includes tracks found in the tracking system
that originate from the primary vertex but are not
associated with any reconstructed object.
 
Depending on the analysis channel, events are selected by lepton triggers~\cite{TRIG-2018-05,TRIG-2018-01}, $\met$ triggers~\cite{TRIG-2019-01} or $b$-jet triggers~\cite{ATLAS:2021piz}. The uncertainty in the combined integrated luminosity is 1.7\%~\cite{ATLAS-CONF-2019-021}, obtained using the LUCID-2 detector~\cite{LUCID2} for the primary luminosity measurements.
Events accepted by lepton and $\met$ triggers are required to meet the standard ATLAS data-quality assessment
criteria~\cite{DAPR-2018-01} to ensure that all subdetector systems were functioning normally. This leads to an integrated luminosity of
3.2~fb$^{-1}$, 33.0~fb$^{-1}$, 44.3~fb$^{-1}$ and 58.5~fb$^{-1}$ in 2015, 2016, 2017 and 2018 data-taking, respectively,
with a total integrated luminosity of 139.0~fb$^{-1}$.
 
Events accepted by $b$-jet triggers are required to meet additional criteria to ensure the validity of the online beam-spot position measurement, which affects the performance of the algorithm used to determine the hard-scatter primary-vertex position.
This leads to the exclusion of the 2015 data and to a reduced integrated luminosity of 24.6~fb$^{-1}$, 43.7~fb$^{-1}$
and 57.7~fb$^{-1}$ in 2016, 2017 and 2018 data-taking, respectively, with a total integrated luminosity of 126.0~fb$^{-1}$.
The $b$-jet trigger chains considered in this paper require at least four jets, among which two jets are $b$-tagged
by the online version of the $b$-tagging boosted decision tree algorithm~\cite{ATL-PHYS-PUB-2016-012}.

Dedicated Monte Carlo (MC) simulated event samples are used to aid in the
estimation of the background from SM processes and to model the
dark matter or invisible Higgs boson signal. All simulated events were processed through a
simulation of the ATLAS detector and its response~\cite{SOFT-2010-01,Agostinelli:2002hh}.
The simulated events are reconstructed with the same algorithms as used for data.
They contain a realistic modelling of additional $pp$ collisions in the same or neighbouring bunch crossings (pile-up), obtained
by overlaying minimum-bias events simulated using the soft QCD processes of \PYTHIA[8.186]~\cite{Sjostrand:2014zea,Sjostrand:2007gs}
with the \NNPDF[2.3lo] set of parton distribution functions (PDFs)~\cite{Ball:2012cx} and the
A3~\cite{ATL-PHYS-PUB-2016-017} set of tuned parameters.
 
Standard Model processes producing one or two top quarks
were modelled using \POWHEGBOX[v2]~\cite{Frixione:2007nw,Nason:2004rx,Frixione:2007vw,Alioli:2010xd}
and normalised to cross sections calculated to next-to-next-to-leading-order (NNLO) plus next-to-next-to-leading-logarithm (NNLL)~\cite{Czakon:2011xx} accuracy in QCD.
Processes involving the production of one (two) vector boson(s) were simulated using \SHERPA[2.2.1]~(2.2.2)~\cite{Gleisberg:2008ta,Gleisberg:2008fv,Cascioli:2011va,Schumann:2007mg,Hoeche:2012yf} and
normalised to cross sections at NNLO~\cite{Catani:2007vq} (NLO) accuracy in QCD. The $t\bar{t}{+}V$ ($V=W,Z$), $tZ$ and $tWZ$ processes were modelled by
\MGNLO[2.3.3]~\cite{Alwall:2014hca} and normalised to NLO cross-section accuracy in QCD~\cite{Alwall:2014hca,deFlorian:2016spz,Alwall:2014hca}.
Lastly, $t\bar{t}H$ background events, with visible Higgs boson decay modes, were generated by \POWHEGBOX[v2].
 
Signal samples for dark matter produced in association with a \ttbar
pair (DM+\ttbar) were generated using a leading-order (LO) matrix element, with up to one
extra parton, in the \MGNLO[2.6.7]~\cite{Alwall:2014hca}
generator interfaced to \PYTHIA[8.244] and using the CKKW-L merging algorithm~\cite{Lonnblad:2001iq}.
The five-flavour scheme \NNPDF[3.0nlo]~\cite{Ball:2014uwa} PDF set was used.
The top quark decay was simulated using \MADSPIN~\cite{Artoisenet:2012st}.
Signal cross sections for this process were calculated to NLO QCD accuracy using the same version of
\MADGRAPH, as suggested in Ref.~\cite{Backovic:2015soa}.
Models with a DM particle mass of 1~\GeV\ and $g_q = g_{\chi} = 1$ are considered, while the masses of the mediators range from 10~\GeV\ to 400~\GeV.
The typical scaling factors from LO to NLO in these calculations range between 1.25 and 1.35,
depending on the mediator mass and whether it is a scalar or a pseudoscalar boson.
Signal samples for dark matter produced in association with a single top
quark were generated using the same settings as for the DM+\ttbar\
samples. For these signal models, the $tW$ and the $tj$ processes were generated separately.
Each one was normalised to the LO cross section predicted by the model and then
the samples were combined. No extra partons were generated from the matrix element in this case.
 
Signal samples modelling Higgs boson production in association with a \ttbar pair \((t\bar{t}H)\) were generated using
the \POWHEGBOX[v2]~\cite{Hartanto:2015uka}
generator at NLO with the \NNPDF[3.0nlo]~\cite{Ball:2014uwa} PDF set.
The \(tWH\) signal samples were produced with \MGNLO[2.6.2] in the five-flavour scheme with the \NNPDF[3.0nnlo] PDF set. The top quark and $W$ boson decays were handled by \MADSPIN to account for spin correlations among the decay products. The overlap of the \(tWH\) process with \(t\bar{t}H\) at NLO was removed by using a diagram removal technique~\cite{Frixione:2008yi,Demartin:2016axk}.
Contributions arising from $tH$ production in the \(t\)-channel and \(s\)-channel are below 0.1\% and are therefore neglected.
In these samples, the Higgs boson decays via $ZZ^*$ into neutrinos, and events are normalised using
the total cross section at NLO QCD and electroweak accuracy recommended by the LHC Higgs Cross
Section Working Group~\cite{deFlorian:2016spz,Demartin:2015uha,Demartin:2016axk}.

For the SM background samples, except those generated using \SHERPA,
the \EVTGEN[1.2.0]~\cite{Lange:2001uf} program was used to simulate
the properties of the $b$- and $c$-hadron decays, and \PYTHIA[8.186] with the A14
tune~\cite{ATL-PHYS-PUB-2014-021} was used for the parton showering,
hadronisation model and underlying event. For the \SHERPA\ generator, the default \SHERPA~\cite{Gleisberg:2008ta,Gleisberg:2008fv,Cascioli:2011va,Schumann:2007mg,Hoeche:2012yf} configuration recommended
by its authors was used.
 
The modelling of the response of the various ATLAS subdetectors was performed using $\GEANT$~\cite{Agostinelli:2002hh}
for all the background MC samples, while all signal MC samples were simulated using a faster simulation based on a parameterisation of the calorimeter response and $\GEANT$ for the other detector systems~\cite{SOFT-2010-01}.


\section{Experimental signatures and analysis strategy}
\label{sec:analysis}

 
Each of the analyses requires the presence of at least one $b$-tagged jet and
\met in the event.
To further suppress events
where the \met\ originates from mismeasurements, the ratio of the \met to
its resolution is used to construct the event's missing transverse
momentum significance~\cite{ATLAS-CONF-2018-038}, $\mathcal{S}$.

The tt0L analysis targets decay chains with no leptons in the final state. The tt0L analysis published
in Ref.~\cite{SUSY-2018-12} focuses on a set of signal regions (SR), referred to as \noLhigh\ in this paper, defined
so as to obtain maximal significance for the targeted signal events. It
uses the $\met$ triggers to select events with large missing transverse momentum
and requires at least one highly
energetic, hadronically decaying top quark candidate. An additional set of SRs,
referred to as \noLlow, is presented for the first time in this paper and extends the previously published results by relying on a combination of $\met$ and $b$-tagged jet triggers to retain events with lower-momentum jets that fail one of the
\noLhigh\ analysis criteria. 
 
The two other analyses considered in this paper target leptonic final states. The \oneL\ analysis~\cite{SUSY-2018-07} selects events with exactly one
lepton ($e$ or $\mu$) and it is based on the $\met$ triggers. The \twoL\ analysis~\cite{SUSY-2018-08} targets events with exactly two opposite-charge leptons ($e$ or $\mu$) in the final states selected with dilepton triggers.

Common event-quality criteria and object reconstruction and identification definitions
for leptons, jets, $b$-tagged jets and \met
are applied in all analyses considered in the combination.
Minimum \pt requirements of $4.5\;\GeV$ and $20\;\GeV$ are applied to leptons and jets, respectively.
The lepton multiplicity requirements guarantee that the 0-lepton, 1-lepton and 2-lepton channels are by construction
non-overlapping.
Further kinematics, isolation and identification requirements are applied to each object in the individual analysis channels and
optimised for each topology. These details, together with the details of the reconstruction algorithms, quality requirements and efficiencies can be found in
Refs.~\cite{SUSY-2018-12,SUSY-2018-07,SUSY-2018-08} and in Appendix~\ref{app:var}.
For all analyses in this paper, background-enriched selections (control regions, CR) are defined so as to allow the data to aid in the
estimation of the dominant SM backgrounds, and validation regions (VR) are
used to verify the robustness of these estimates.
A dedicated background estimation strategy was developed for each channel~\cite{SUSY-2018-12,SUSY-2018-07,SUSY-2018-08},
using independent control regions for all dominant SM processes.
 
Observed and expected
event yields in the signal and control regions are used in Poisson
probability functions to build
likelihood functions, which are combined in a profile likelihood fit. A
profile likelihood ratio is employed in the $\mathrm{CL_s}$ method to exclude at 95\% CL the
signal-plus-background hypothesis for the signal models considered~\cite{Baak:2014wma,Cowan:2010js,Junk:1999kv,Read:2002hq}.

\subsection{Description of analysis channels}
\label{sec:analyses_detail}

\paragraph{The \noL\ analysis}
 
\begin{figure}[p]
\begin{center}
\begin{subfigure}[t]{0.96\textwidth}
\centering
\includegraphics[width=0.99\textwidth]{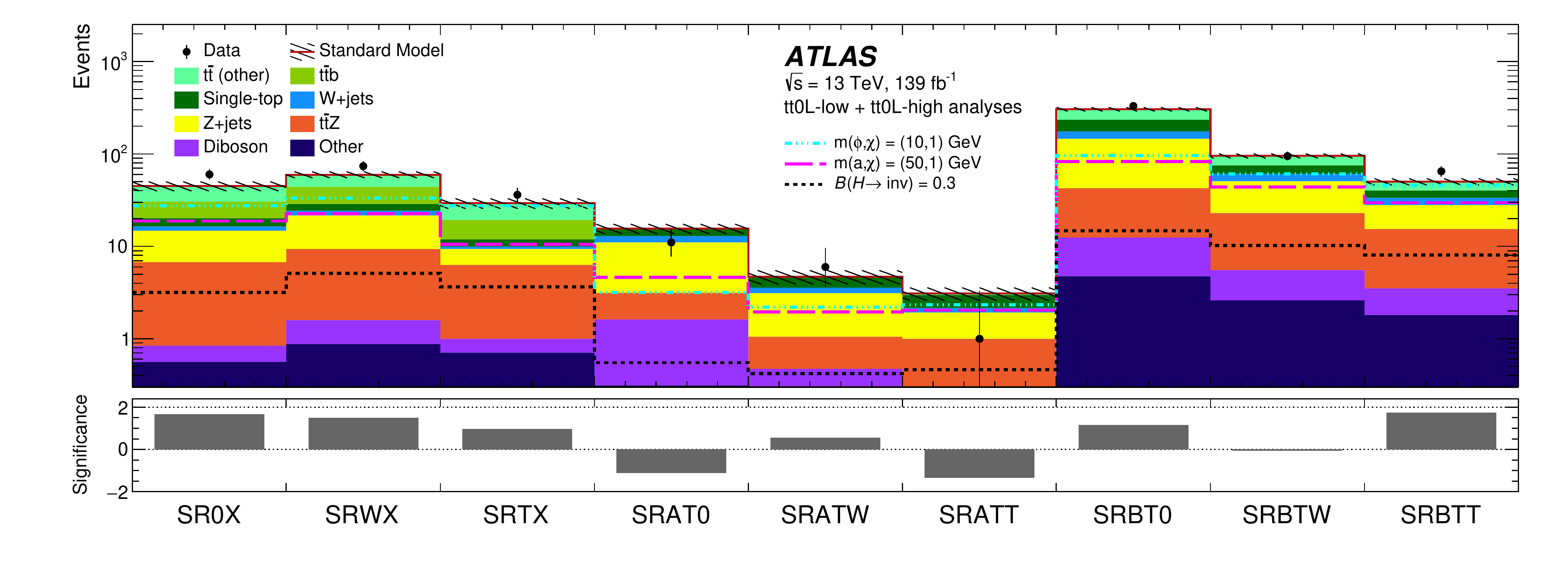}
\caption{}
\label{fig:srplots_0L}
\end{subfigure}
 
\begin{subfigure}[t]{0.48\textwidth}
\centering
\includegraphics[width=0.99\textwidth]{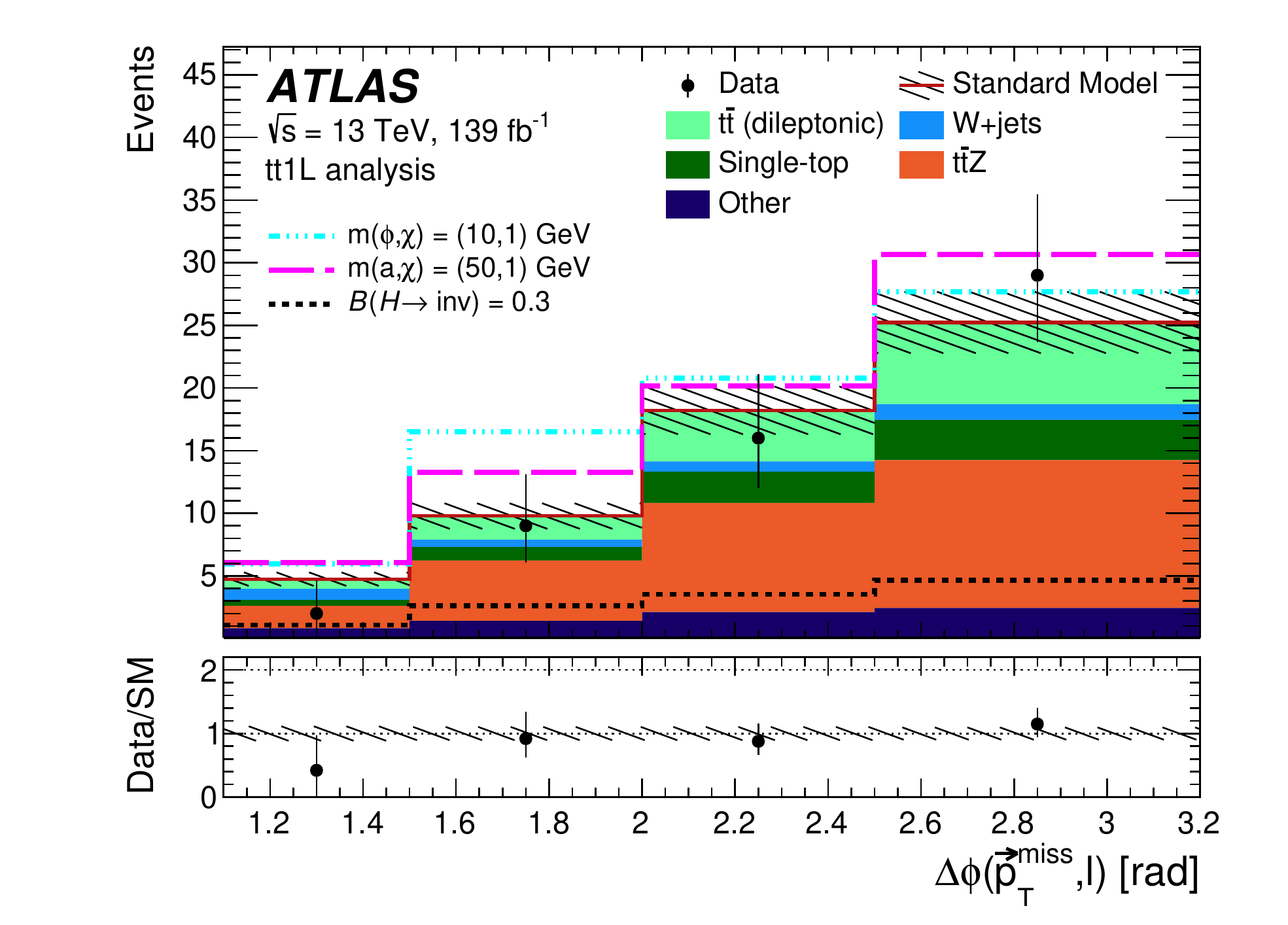}
\caption{}
\label{fig:srplots_1L}
\end{subfigure}
\begin{subfigure}[t]{0.48\textwidth}
\centering
\includegraphics[width=0.99\textwidth]{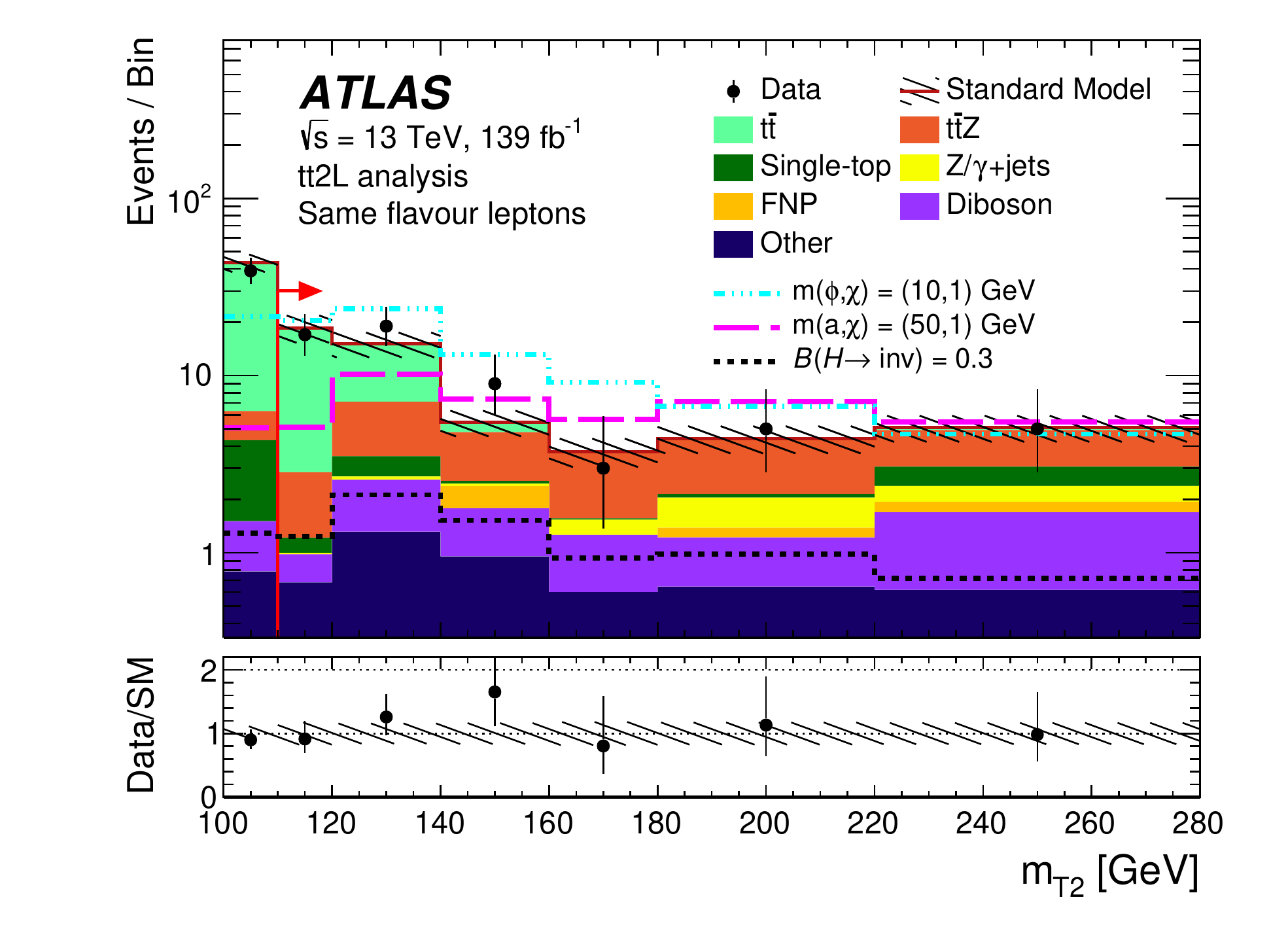}
\caption{}
\label{fig:srplots_2L}
\end{subfigure}
\caption{(a) Post-fit signal region yields for the \noLhigh\ and the \noLlow\ analyses.
The bottom panel shows the statistical significance~\cite{ATL-PHYS-PUB-2020-025} of the difference
between the SM prediction and the observed data in each region.
The definitions of SRAT0, SRATW, SRATT, SRBT0, SRBTW and SRBTT can be found in Ref.~\cite{SUSY-2018-12}.
Representative post-fit distributions are presented for (b) the \oneL\ and (c) the \twoL\ analyses: each bin of these distributions,
starting from the red arrow in (c), corresponds to a single SR included in the fit.
In the \noLlow\ analysis, `\ttbar\ (other)' represents \ttbar\ events without extra jets or events with extra light-flavour jets.
In the \twoL\ analysis, `FNP' includes the contribution from fake/non-prompt lepton background arising from jets
(mainly $\pi/K$, heavy-flavour hadron decays and photon conversion) misidentified as leptons, estimated in a purely data-driven way.
`Other' includes contributions from $\ttbar{+}W$, $tZ$ and $tWZ$ processes, and also \ttbar (semileptonic) for the \oneL\ analysis.
The total uncertainty in the SM expectation is represented with hatched bands and the expected distributions
for selected signal models are shown as dashed lines. }
\end{center}
\end{figure}

The experimental signature targeted in this channel consists of at least four jets, two of which are \btagged, and large missing transverse momentum ($\met > 160\;\GeV$).  Events with electrons, muons or $\tau$-leptons are rejected.
The \noLhigh\ selection consists of events with high missing transverse momentum (\met trigger, $\met > 250\;\GeV, \mathcal{S}> 14$) where at least one of the reconstructed large-radius jets ($R=1.2$) has a mass consistent with one produced from a boosted hadronic top quark decay.
This selection corresponds to signal region selections SRA and SRB in Ref.~\cite{SUSY-2018-12}, which were originally optimised to be sensitive to
high-mass supersymmetric partners of the top quark and are divided into three categories
(TT, TW and T0), depending on whether the subleading large-radius jet ($R=1.2$)
has an invariant mass consistent with a top quark, a $W$ boson or
neither. Given the signature and kinematic similarity between the signal considered in this paper and
top squark pairs decaying into a top quark and a neutralino, these regions can also be used to
constrain DM+$\ttbar$ models.
SRA and SRB are orthogonal because of a requirement on the
$\chi^2$-based stransverse mass variable, $m_{\mathrm T2, \chi^2}$, and are statistically combined.
This stransverse mass~\cite{Lester:1999tx,Barr:2003rg} variable is designed to reconstruct the
mass of two heavy particles produced in an event and decaying symmetrically into
an invisible particle and a top quark, as expected in supersymmetric top quark topologies. It relies on a $\chi^2$-based method to identify the hadronically decaying top
quark candidates.
The background estimation in SRA and SRB is aided by means of
dedicated control regions for all dominant SM
processes: $t\bar{t}Z$, $Z$+jets, \ttbar, single top quark in the $tW$ channel, and $W$+jets. Event
yields with updated jet calibrations~\cite{JETM-2018-05} were recalculated with
respect to Ref.~\cite{SUSY-2018-12} and are presented in
Figure~\ref{fig:srplots_0L}. The yield change due to the new calibration is between
6\% and 15\% and is reflected in the signal predictions as well. This is because the new calibration decreases the
contribution of events with artificially increased missing transverse momentum, due to mismeasured jets in the analysis, by reducing the
tails of the  missing transverse momentum significance distribution.
This translates to a slightly higher signal-to-background ratio in the
\noLhigh\ SRs.
 
The \noLlow\ selection is newly added in this paper with the aim of improving the sensitivity of the 0-lepton channel
for dark matter models by selecting final states with lower missing transverse momentum
and/or lower-momentum objects.
The \noLlow\ selection is therefore designed to maximise the sensitivity to
DM+\ttbar signals with low mediator masses
($m(\phi),m(a)<100~\GeV$). Events are selected by a combination of $\met$ triggers and $b$-jet triggers.
Events selected with $\met$ triggers must fulfil $\met > 250\;\GeV$
to ensure the triggers are fully efficient. In addition, they are required to have either no large-radius jets consistent with highly energetic top quark candidates or  $\mathcal{S}$  bounded between 8 and 14, to ensure orthogonality with the \noLhigh\ selections.
Events selected with $b$-jet triggers must fulfil the requirement that the missing transverse momentum be between 160 and 250~\GeV.
The lower bound suppresses the multi-jet background contamination, while
the upper bound ensures orthogonality with \noLhigh. The online $b$-tagged jet
candidates must match the offline $b$-tagged jet candidates within a cone of $\Delta R = 0.2$.
 
Three signal regions are defined, SR0X,
SRWX and SRTX, according to the mass of the heaviest large-radius jet, which is used to infer the \pt\ of the most energetic
top quark in the event. SR0X requires
that no large-radius jets (0) are present, while in SRWX and SRTX the
mass of the heaviest large-radius jet has to be respectively lower
or higher than 130~\GeV, i.e.\ in the neighbourhood of the $W$ boson
(W) or top quark (T) mass.
 
An optimal categorisation of the events in which a top quark is produced
is particularly important in the \noLlow\ analysis because of
the high top quark background rate that varies significantly with respect to the
top quark transverse momentum. In order to maximise the correlation of the large-radius jet mass
with the \pt of the most energetic top quark,  jet reconstruction in the \noLlow\ analysis uses a
smaller radius parameter value ($R=1.0$) than in the \noLhigh\ analysis.
This is  associated with newly
introduced lower selections on the large-radius jet transverse
momentum ($\pt>200\;\GeV$), and invariant mass
($m_{\textrm{large-radius jet}} > 40\;\GeV$).
A requirement rejecting high-mass, large-radius jets with
radius parameter $R=1.2$, as defined in the \noLhigh\ analysis, is also applied to ensure orthogonality between the two analyses.

The final states targeted in the \noLlow\ analysis are less energetic than in the \noLhigh\ analysis, such that the decay
products of the top quarks are expected to be less energetic. Hence no subleading large-radius jet is required (X).
 
\begin{figure}[tbp]
\begin{center}
\begin{subfigure}[t]{0.48\textwidth}
\centering
\includegraphics[width=0.99\textwidth]{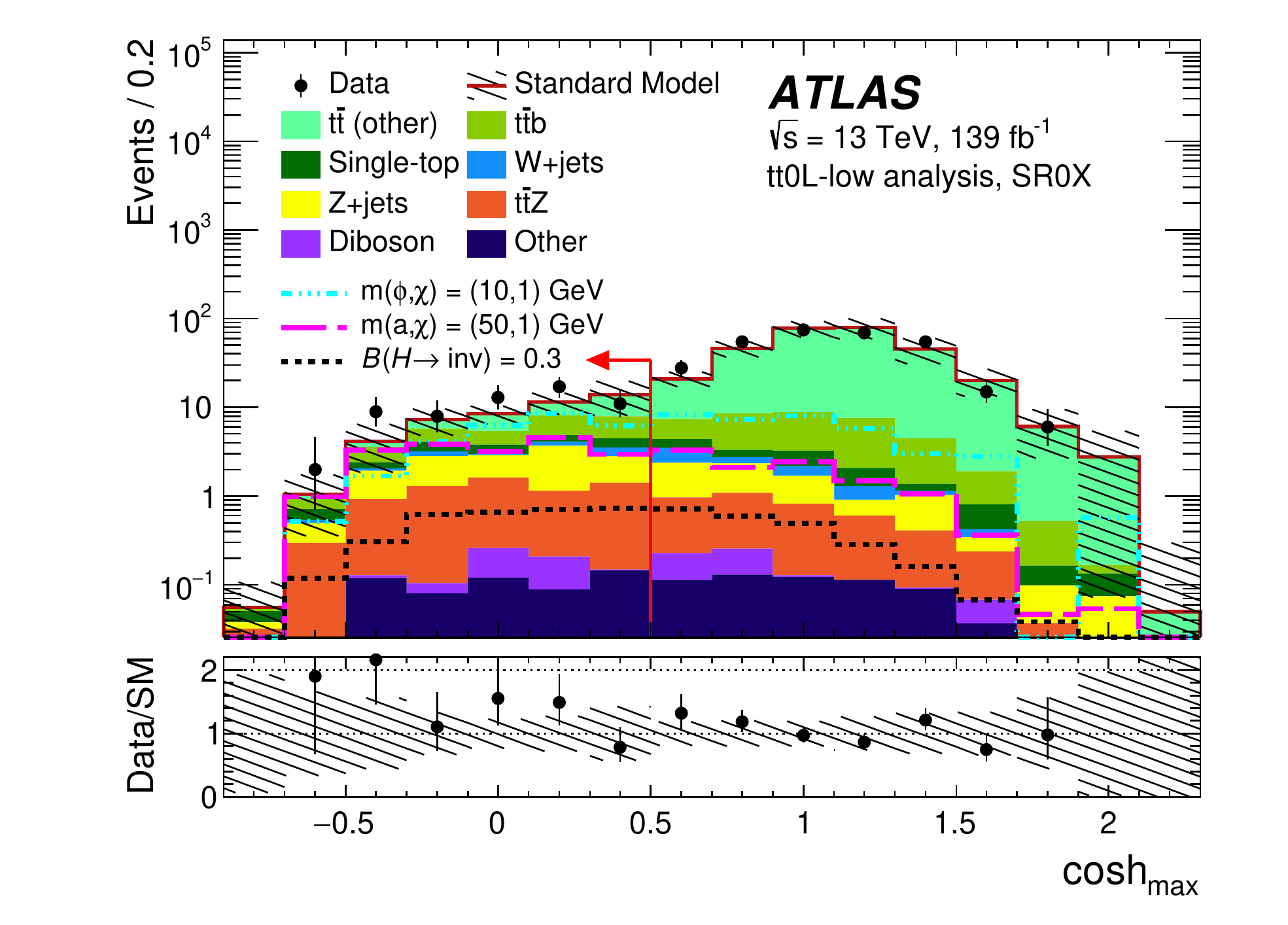}
\caption{}
\end{subfigure}
\begin{subfigure}[t]{0.48\textwidth}
\centering
\includegraphics[width=0.99\textwidth]{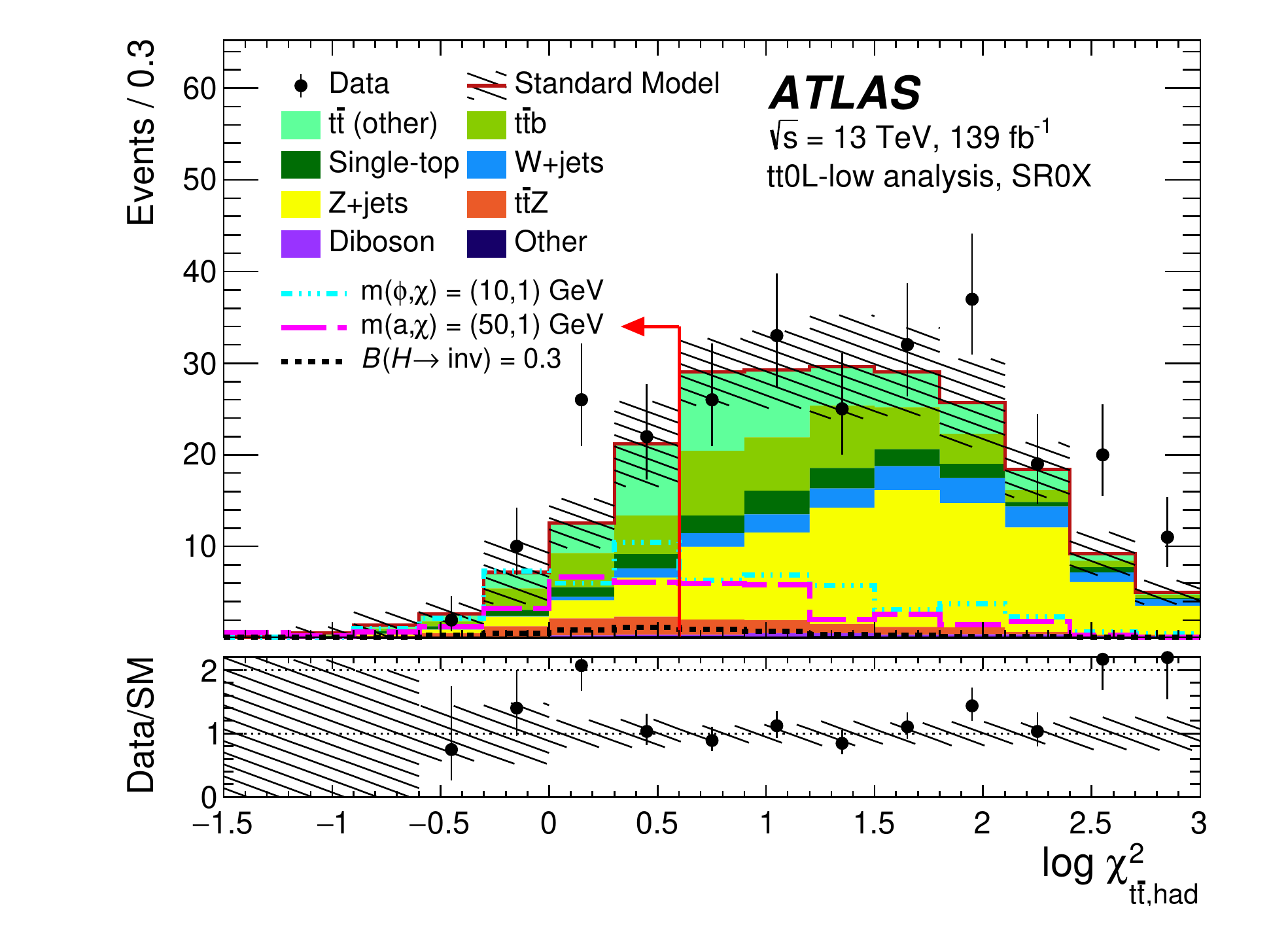}
\caption{}
\end{subfigure}
\caption{Distributions of (a) \coshmax and (b) \chitwotop in SR0X events passing all the SR requirements
except those on the variable being presented (which are indicated by the arrows).
The contributions from all SM backgrounds are shown after the simultaneous profile likelihood  fit to all \noLlow\ CRs,
with the hatched bands representing the total uncertainty.
The category `\ttbar\ (other)' represents \ttbar\ events without extra jets or events with extra light-flavour jets.
`Other' includes contributions from $\ttbar{+}W$, $tZ$ and $tWZ$ processes.
The expected distributions for selected signal models are shown as dashed lines.
The underflow (overflow) events are included in the first (last) bin.
The bottom panels show the ratio of the observed data to the total SM background prediction,
with the hatched area representing the total uncertainty in the background prediction.}
\label{fig:key_discriminants}
\end{center}
\end{figure}

The full list of requirements for the three \noLlow\ signal regions is reported in
Table~\ref{tab:lowmet_sr}.
Three angular separation variables are used: $\Delta\phi_{\min}(\vec{p}_{\text{T,1-4}},\vec{p}_{\text{T}}^{\text{miss}})$
to reduce the contamination from multi-jet events,
\drbb\ to reduce the contamination from $b$-jets from gluon splitting as present in $Z$+jets events,
and $\Delta R_{\min}$(large-radius jet, $b$-tagged jets) to enhance the fraction of events with hadronic top quark decays, as in the signal.
The SM background originating from the top quark is reduced by using the \coshmax\ variable. This variable aims to identify events
where a leptonically decaying $W$ boson is the source of all \met\ in the event because the lepton is not reconstructed.
The hyperbolic cosine of the pseudorapidity difference between the missed $W$ boson candidate and each of the two $b$-tagged jets selected in the event
is estimated, and the \coshmax\ variable is defined to be the larger of the two values:
\begin{align}
\coshmax = \max\{\cosh\Delta\eta^{1}_{W,b}, \cosh\Delta\eta^{2}_{W,b}\}. \nonumber
\end{align}
The value of $\cosh\Delta\eta_{W,b}$
is estimated by solving the kinematics of the top quark decay,
assuming $m_{W} \ll \pt^{W}\cdot\cosh\eta_{W} \sim \met\cdot\cosh\eta_{W}$, given that $\met > 160~\GeV$ and $\cosh\eta_{W}\ge1$.
For events with \ttbar\ pairs decaying fully hadronically, the approximation $\pt^{W} \sim \met$ is not valid, resulting in unphysical values, e.g.\ $\coshmax < 1$,  while events with high \coshmax\ values are likely to contain a top quark decaying leptonically and are excluded from the signal regions.
Since the signal events have two hadronically decaying top quarks, while most of the backgrounds have one or none, a $\chi^2$-based variable, \chitwotop,
is used to identify events with a hadronically decaying top quark pair (Appendix~\ref{app:var}).
The \chitwotop\ variable is constructed using up to six jets in the
event and using the $b$-tagging classification when appropriate to
reconstruct the two $W$ bosons and the two top quarks in the
event. Invariant mass constraint terms for  these particles are
used to build the $\chi^2$. All possible jet assignments are tested and the one with the lowest $\chi^2$ is chosen.
The event distributions for the \coshmax\ and the \chitwotop\ variables are each shown in Figure~\ref{fig:key_discriminants}
after applying all the SR0X selection requirements except those on the variable presented in the distribution.
The transverse momentum of the \ttbar system ($\pt^{\ttbar}$) constructed with the \chitwotop\ method is compared with the \met.
In signal events these two quantities are expected to have similar values, so events with $\pt^{\ttbar}/\met$ values outside a window containing unity are excluded.
 
\begin{table}[htbp]
\centering
\scriptsize
\caption{Selection criteria for the signal regions used in the \noLlow\ analysis.}
\label{tab:lowmet_sr}
 
\begin{tabular}{|>{\centering}p{0.3\textwidth-2\tabcolsep}|>{\centering}p{0.23\textwidth-2\tabcolsep}|>{\centering}p{0.23\textwidth-2\tabcolsep}|>{\centering\arraybackslash}p{0.23\textwidth-2\tabcolsep}|}
\hline
\textbf{Variables} & \textbf{SR0X} & \textbf{SRWX} & \textbf{SRTX} \\
\hline
$N_{\textrm{lepton}}$ & \multicolumn{3}{c|}{$=$ 0} \\
\hline
Orthogonalisation & \multicolumn{3}{c|}{$\met < 250~\GeV$ \hspace{0.01\textwidth} or \hspace{0.01\textwidth} $\mathcal{S} < 14$ \hspace{0.01\textwidth} or \hspace{0.01\textwidth} $m_{\textrm{large-radius jet}}^{R=1.2} < 120~\GeV$} \\
\hline
\multirow{2}{*}[0.6mm]{\met[GeV]} & \multicolumn{3}{c|}{$>$ 160} \vspace{-1.2mm}\\
& \multicolumn{3}{c|}{$<$ 250, when passing $b$-jet triggers} \\
\hline
$\mathcal{S}$ & \multicolumn{3}{c|}{$>$ 10} \\
\hline
$\Delta\phi_{\min}(\vec{p}_{\text{T,1-4}},\vec{p}_{\text{T}}^{\text{miss}})$ & $>$ 1.0 & \multicolumn{2}{c|}{$>$ 0.5} \\
\hline
\drbb & \multicolumn{3}{c|}{$>$ 1.2} \\
\hline
$N_{\textrm{large-radius jet}}$ & $=$ 0 & \multicolumn{2}{c|}{$>$ 0} \\
\hline
$m_{\textrm{large-radius jet}}$ [GeV] & --- & (40, 130) & $\geq$ 130 \\
\hline
$\Delta R_{\min}$(large-radius jet, $b$-tagged jets) & \multicolumn{2}{c|}{---} & $<$ 1.2 \\
\hline
\coshmax & $<$ 0.5 & $<$ 0.6 & $<$ 0.7 \\
\hline
\chitwotop & $<$ 4 & $<$ 6 & $<$ 8 \\
\hline
$\pt^{\ttbar}/\met$ & (0.7, 1.2) & \multicolumn{2}{c|}{(0.5, 1.2)} \\
\hline
\end{tabular}
\end{table}

The main contribution to the SM background in the signal regions originates from single-top-quark events in the $tW$ channel and  \ttbar\ events,
with a lepton missed by the reconstruction algorithms,
and $Z\to\nu\nu$ events ($Z$+jets, $\ttbar{+}Z$). Unlike in the \noLhigh\ SRs, \ttbar\ is the dominant contribution in the \noLlow\ selections, due to the lower \met\ requirements.
Dedicated control regions are used to aid the background estimation in the signal regions for all
dominant processes. These CRs are similar to the ones in the  \noLhigh\ analysis for single top quarks in the $tW$ channel, $Z$+jets and $\ttbar{+}Z$.
No control regions are defined for $W$+jets since it represents less than 5\% of the total background in all three \noLlow\ signal regions.
Due to the importance of the \ttbar\ background in the \noLlow\ selections, $\ttbar$ events with extra $b$-hadrons ($\ttbar{+}b$) are treated separately from \ttbar\ events without extra jets or events with extra light-flavour jets (referred to as \textit{other}) in the \noLlow\ analysis, and distinct control regions are defined for each of the two components. This is done because $\ttbar\text{+}b$ events are more likely to pass the signal region selections than \ttbar (other) events because of the presence of extra $b$-jets, which makes it more difficult to isolate and reconstruct the \ttbar\ system. In practice, the \ttbar\ (other) and $\ttbar{+}b$ event distributions exhibit a significant shape difference in the \coshmax\ variable. Moreover, the observed simulation mismodelling for $\ttbar{+}b$ events is larger than for \ttbar\ (other) events and can be corrected for by separating the two components.

The background estimates are validated in dedicated, non-overlapping,
validation regions, which require zero leptons and are orthogonal to the signal region selections.
In these regions, the background prediction agrees with the data to within one standard deviation ($1\sigma$).
More details are given in Appendix~\ref{app:bkg}.
 
The expected numbers of events estimated in a simultaneous profile likelihood fit to all \noLlow\ CRs
are shown in Table~\ref{tab:lowmetyield}. The observed data are compatible with the prediction, agreeing to within $2\sigma$ in each signal region.
 
The results presented in this paper show the final combination of the \noLlow\ and \noLhigh\ analyses, estimated in a simultaneous fit of all CRs and SRs.
The details of this combination and the single-channel individual limits are discussed in Appendix~\ref{sec:lowmetresults}.
 
\begin{table}[htbp]
\centering
\scriptsize
\caption{Expected and observed numbers of events in SR0X, SRWX and SRTX. The background
yields and uncertainties are shown after the simultaneous profile likelihood  fit to all \noLlow\ CRs.
The category `\ttbar\ (other)' represents \ttbar\ events without extra jets or events with extra light-flavour jets.
`Other' includes contributions from $\ttbar{+}W$, $tZ$ and $tWZ$ processes.
The quoted background uncertainties include both the statistical and systematic contributions, while the signal uncertainties are purely statistical.}
\label{tab:lowmetyield}
\begin{tabular}{p{0.26\textwidth-2\tabcolsep}>{\centering}p{0.25\textwidth-2\tabcolsep}>{\centering}p{0.25\textwidth-2\tabcolsep}>{\centering\arraybackslash}p{0.25\textwidth-2\tabcolsep}}
\hline
\textbf{Process} & \textbf{SR0X} & \textbf{SRWX} & \textbf{SRTX} \\
\hline
Observed data & 60\leavevmode\phantom{.00 $\pm$ 0.00} & 74\leavevmode\phantom{.00 $\pm$ 0.00} & 36\leavevmode\phantom{.00 $\pm$ 0.00} \\
\hline
Expected SM events & 45\bblank\ $\pm$ 8\bblank & 59\bblank\ $\pm$ 6\bblank & 28\bblank\ $\pm$ 5\bblank \\
\hline
\ttbar\ (other) & 14\bblank\ $\pm$ 4\bblank & 15\bblank\  $\pm$ 4\bblank  & 9.4\blank\ $\pm$ 3.5  \\
$\ttbar{+}b$      & 10\bblank\ $\pm$ 7\bblank & 15.0\blank\ $\pm$ 3.1\blank & 7.2\blank\ $\pm$ 2.8  \\
Single-top      & 3.8\blank\ $\pm$ 3.0      & 4.3\blank\  $\pm$ 2.6       & 1.9\blank\ $\pm$ 1.5  \\
$Z$+jets        & 8.0\blank\ $\pm$ 1.6      & 12.1\blank\ $\pm$ 2.3\blank & 3.1\blank\ $\pm$ 0.8  \\
$W$+jets        & 1.6\blank\ $\pm$ 1.1      & 2.7\blank\  $\pm$ 2.1       & 0.6\blank\ $\pm$ 0.6  \\
$\ttbar{+}Z$      & 5.9\blank\ $\pm$ 1.0      & 7.8\blank\  $\pm$ 1.3       & 5.3\blank\ $\pm$ 1.1  \\
Diboson         & \blank0.28 $\pm$ 0.20     & 0.7\blank\  $\pm$ 0.4       & \blank0.30 $\pm$ 0.19 \\
Other           & \blank0.55 $\pm$ 0.15     & \blank0.88  $\pm$ 0.24      & \blank0.70 $\pm$ 0.22 \\
\hline
Pre-fit \ttbar      & 15\leavevmode\phantom{.00 $\pm$ 0.00} & 17\leavevmode\phantom{.00 $\pm$ 0.00} & 9.8\leavevmode\phantom{0 $\pm$ 0.0}   \\
Pre-fit $\ttbar{+}b$  & 7\leavevmode\phantom{.00 $\pm$ 0.0}   & 11.5\leavevmode\phantom{0 $\pm$ 0.00} & 5.6\leavevmode\phantom{0 $\pm$ 0.0}   \\
Pre-fit Single-top  & 7.1\leavevmode\phantom{0 $\pm$ 0.0}   & 8.2\leavevmode\phantom{0 $\pm$ 0.0}   & 3.6\leavevmode\phantom{0 $\pm$ 0.0}   \\
Pre-fit $Z$+jets    & 6.1\leavevmode\phantom{0 $\pm$ 0.0}   & 9.2\leavevmode\phantom{0 $\pm$ 0.0}   & 2.3\leavevmode\phantom{0 $\pm$ 0.0}   \\
Pre-fit $\ttbar{+}Z$  & 5.9\leavevmode\phantom{0 $\pm$ 0.0}   & 7.9\leavevmode\phantom{0 $\pm$ 0.0}   & 5.4\leavevmode\phantom{0 $\pm$ 0.0}   \\
\hline
Benchmark signal models \\
\hline
DM $m(\phi, \chi) = (10, 1)\,\GeV$ & 27.4\blank\ $\pm$ 2.4\blank & 33.2\blank\ $\pm$ 2.2\blank & 27.5\blank\ $\pm$ 2.2\blank \\
DM $m(a, \chi) = (50, 1)\,\GeV$    & 18.8\blank\ $\pm$ 1.3\blank & 22.6\blank\ $\pm$ 1.5\blank & 10.6\blank\ $\pm$ 1.0\blank \\
$H \rightarrow$ inv  ($\br=100$\%)              & 10.52       $\pm$ 0.34      & 17.1\blank\ $\pm$ 0.4\blank & 12.1\blank\ $\pm$ 0.4\blank \\
\hline
\end{tabular}
\end{table}

\paragraph{The \oneL\ analysis}

This analysis requires exactly one lepton ($e$ or $\mu$), at least
four jets, two of which must be \btagged, and $\met > 230\;\GeV$, and was
designed to target spin-0 DM models.
The \MET significance $\mathcal{S}$ must be above 15 and, only for this
analysis, it considers only jets and leptons in the
events and their resolution, as described in Ref.~\cite{SUSY-2013-15}.
A recursive variable-radius reclustering algorithm~\cite{SUSY-2016-16} is applied to the jets
to identify at least one large-variable-radius jet loosely consistent with a
top quark ($m_{\mathrm{top}^{\mathrm{reclustered}}} > 150\;\GeV$). The use of a variable-radius algorithm, instead of a
fixed-radius one, increases the acceptance of both highly boosted events and less boosted events when no explicit categorisation is
performed.
In addition, a requirement on the `topness' likelihood variable~\cite{Graesser:2012qy} is used to distinguish between the signal and dileptonic decays in SM \ttbar\ events where one of the leptons is misidentified or outside the acceptance. This variable quantifies how well each event satisfies the dileptonic \ttbar\ hypothesis, using the top quark and $W$ boson mass constraints and a requirement that the centre-of-mass energy of the event is minimised.
The \MET\ triggers were used to select
data that then populate this SR.  This region is
divided into four disjoint regions according to the azimuthal distance between
the \met and the lepton momentum,
$\Delta\phi(\vec{p}_{\textrm{T}}^{\textrm{miss}},\ell)$, which is presented in Figure~\ref{fig:srplots_1L}
and is found to be larger for pseudoscalar mediator models. The binning also maximises the sensitivity for
scalar mediator models, which are more similar to the
background but are characterised by a larger production cross section at low masses.
An additional requirement of
$\Delta\phi(\vec{p}_{\textrm{T}}^{\textrm{miss}},\ell) > 1.1$ is applied to suppress the SM background.
The dominant backgrounds, $t\bar{t}$ and $t\bar{t}Z$, are estimated
by means of dedicated CRs.
 
\paragraph{The \twoL\ analysis}
 
The last analysis considers events with two opposite-charge leptons
($e$ or $\mu$), at least one \btagged jet and large values of  \met  significance ($\mathcal{S} > 12$), exploiting events
collected with dilepton triggers.
Events are then separated into two categories depending on whether the two leptons have the same or
different flavour, and  in the same-flavour selection an additional
requirement of $|m_{\ell\ell}-m_Z|>20$~\GeV~is added to suppress
the \(Z\)+jets background. In this selection, the main discriminating
variable is the leptonic stransverse mass
$m_{\textrm{T2}}$~\cite{Barr:2003rg,Lester:1999tx}, which is used to
bound the individual masses of a pair of identical particles that are each
presumed to have decayed into one visible and one invisible
particle. This quantitiy is used to bound dileptonic top pair decays. To maximise the search sensitivity, the $m_{\textrm{T2}}$
spectrum is divided into six bins, starting from 110~\GeV. The
$m_{\textrm{T2}}$ distribution for selected events with two leptons
with the same flavour is presented in Figure~\ref{fig:srplots_2L}.
In this search, the main backgrounds are from $t\bar{t}$, $t\bar{t}Z$,
single-top-quark $tW$-channel, $Z$+jets, and diboson
processes. These backgrounds are estimated with MC simulations
and normalised with data in orthogonal CRs for
the dominant contributions ($t\bar{t}$ and $t\bar{t}Z$), while
the background arising from fake/non-prompt leptons is estimated
directly from the data.
 
\subsection{Orthogonalisation}
 
In order to combine the results of the different searches,
the searches are required to be statistically independent and
any possible overlaps of kinematic regions were investigated and removed as described in the following.
The three analysis channels are disjoint because of their requirements on
lepton multiplicity. The \noLhigh\ and \noLlow\ channels
are kept orthogonal by the requirements on the large-radius jet as
well as on the \met and its significance, $\mathcal{S}$. In addition, one of the $Z$+jets CRs
in the \noLhigh\ analysis, denoted by CRZAB-T0 in Ref.~\cite{SUSY-2018-12},
is not considered and a single control region, CRZAB-TTTW, is used to normalise
the $Z$+jets process in all SRs of the \noLhigh\ analysis. This has negligible
impact on the \noLhigh\ analysis results and it is done
to ensure orthogonality between the $Z$+jets CRs in the \noLhigh\ and \noLlow\
analyses, as those events are used to normalise the $Z$+jets background in the \noLlow\ analysis. To the same end,
the $Z$+jets CR in the \noLlow\ analysis only selects events with either
$N_{\textrm{large-radius jet}} < 2$ or subleading large-radius
jet mass~$< 60$~\GeV.
 
The CRs used to normalise the $t\bar{t}Z$ background overlap.
The three analysis channels share a common strategy to determine the amount of
$t\bar{t}Z$ (with \(\Zboson\rightarrow\nu\nu\)) background in their SRs.
The strategy is to construct CRs requiring three charged leptons in order to maximise their $t\bar{t}Z$ (with \(\Zboson\rightarrow\ell\ell\))
event content, which once determined can be scaled by the ratio of
\(\Zboson\rightarrow\nu\nu\) to \(\Zboson\rightarrow\ell\ell\) branching fractions.
These control regions differ only in minor selections adapted to the SR of each specific channel.
In the combination, the $t\bar{t}Z$ estimation is harmonised by using the
most inclusive CR$_{\ttbar Z}$, from the \twoL\ analysis~\cite{SUSY-2018-08}, as
a common CR across all channels. The fitted normalisation parameter value
obtained in the combination is consistent within 1\% with the one published in Ref.~\cite{SUSY-2018-07}.


\FloatBarrier
\section{Statistical combination and uncertainties}
\label{sec:comb}


The statistical combination of the analyses considered in this paper
consists of maximising a profile likelihood
ratio~\cite{Cowan:2010js} constructed from the product of the
individual analysis likelihoods:
 
\[
\Lambda(\alpha;\theta) = \frac{L\left(\alpha,\hat{\hat{\theta}}(\alpha)\right)}{L\left(\hat{\alpha},\hat{\theta}\right)}.
\]

The \(\alpha\) and \(\theta\) parameters represent, respectively, the parameter of interest
and the nuisance parameters. In the numerator, the nuisance parameters
are set to their profiled values \(\hat{\hat{\theta}}(\alpha)\), which
maximise the likelihood function for fixed values of the parameter of
interest \(\alpha\). In the denominator, both the parameter of
interest and the nuisance parameters are set to the values that
jointly maximise the likelihood: \(\hat{\alpha}\) and
\(\hat{\theta}\), respectively.

For the DM signal model interpretations, upper limits on the signal cross section are calculated following
the $\mathrm{CL_s}$ formalism, using the profile likelihood
ratio as a test statistic.
The parameter of interest is the overall signal strength, defined as a scale factor
multiplying the cross section predicted by the signal hypothesis, and it is bounded from below by zero.
The final result is provided as a ratio of the lowest excluded
signal cross section to the predicted cross section with all couplings set to unity.
For the \hinv\ signal model interpretation, the branching fraction \brinv is considered as the parameter of interest \(\alpha\), following the
implementation described in Refs.~\cite{Verkerke:2003ir,Moneta:2010pm}.
 
As described in Section~\ref{sec:analysis}, for each channel the estimation
of the dominant SM backgrounds is aided by means of dedicated control regions
that constrain free-floating normalisation factors for each
of these backgrounds.

Systematic
uncertainties are modelled in the likelihood function as nuisance
parameters $\theta$ constrained by Gaussian or log-normal probability
density functions~\cite{HIGG-2015-03}.

Three types of
sources of systematic uncertainty are considered: detector-related (experimental)
uncertainties, uncertainties related to the modelling of SM background
processes, and uncertainties related to the modelling of the signal
processes. Regarding the experimental and SM modelling uncertainties,
all details are given in Refs.~\cite{SUSY-2018-12,SUSY-2018-07,SUSY-2018-08} respectively
for the zero-, one- and two-lepton channels. The \noLlow\ channel includes the same
uncertainties as the \noLhigh\ channel and, in addition, uncertainties associated with the
$b$-jet trigger efficiencies. The typical size of these uncertainties is a few percent.
All analyses use common event-quality criteria and object
reconstruction and identification definitions. For this reason, all
experimental systematic uncertainties are treated as correlated across
channels in the statistical combination. The dominant sources of
experimental systematic uncertainty in the combination are the uncertainties
related to the jet energy scale and resolution, followed by either flavour-tagging
uncertainties or uncertainties related to the missing transverse momentum, depending on the analysis channel.

\begin{figure}
\centering
\includegraphics[width=.9\textwidth]{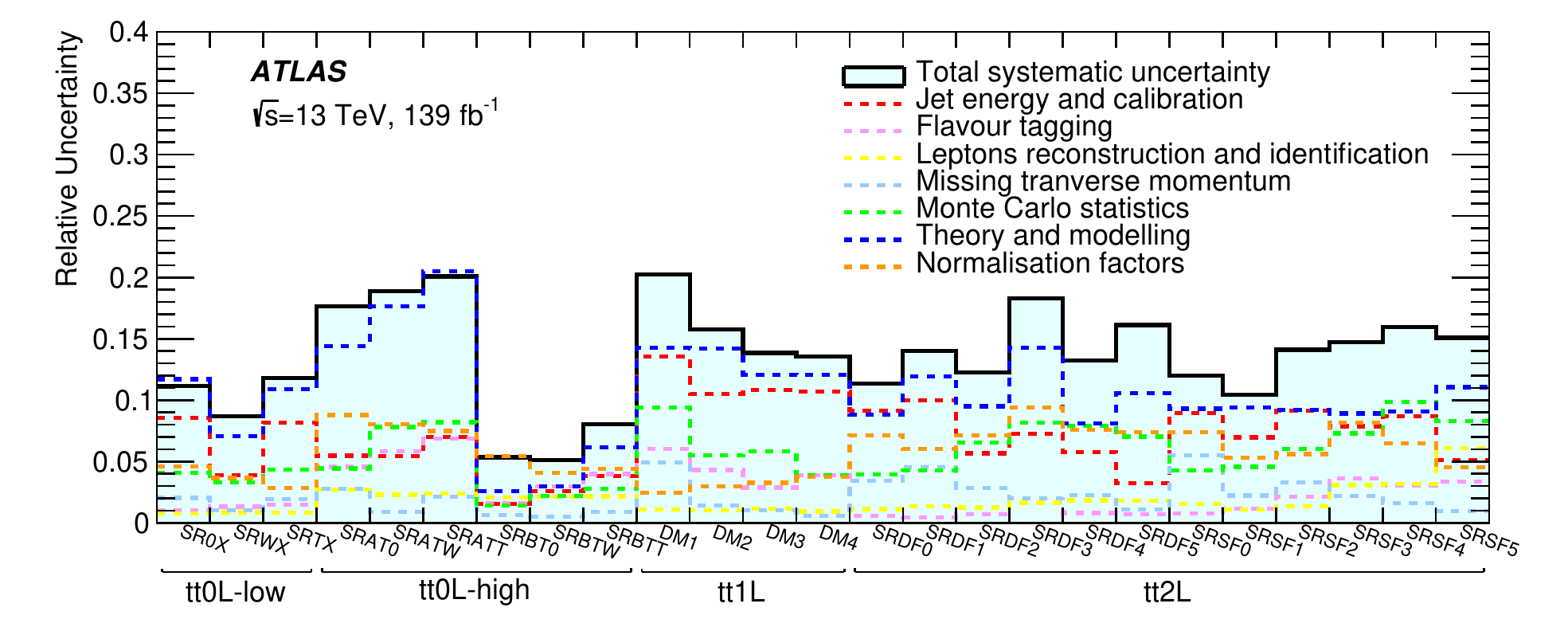}
\caption{Summary of the total uncertainty in the background prediction for each SR
of the \noLlow, \noLhigh, \oneL, and \twoL\ analysis channels in the statistical combination after the profiled likelihood fit. Their dominant contributions are indicated by individual lines. Individual uncertainties can be correlated, and do not necessarily add up in quadrature to the total background uncertainty.}
\label{g:SummUncertainties}
\end{figure}
 
Uncertainties in the modelling of the SM background processes in MC
simulation and their theoretical cross-section uncertainties are also
taken into account. All modelling uncertainties are treated as
uncorrelated across different channels as they probe different regions
of the available phase space.
 
Uncertainties related to the MC modelling of the DM signals include fragmentation and renormalisation scale uncertainties, and the uncertainties
related to the modelling of the parton shower. The impact
of these uncertainties varies from 10\% to 25\%. Uncertainties related to the $t\bar{t}H$ with \hinv\ signal modelling also include
fragmentation and renormalisation scale uncertainties, parton shower uncertainties and PDF uncertainties.
Among these, scale uncertainty effects, which are evaluated in the simplified template cross-section formalism~\cite{deFlorian:2016spz,HIGG-2020-23},
are the dominant contribution and range between 7\% and 17\%.
Signal modelling uncertainties are treated as fully correlated across analysis channels.
 
All sources of uncertainty in the SM backgrounds are summarised in Figure~\ref{g:SummUncertainties}.
In most of the SRs, the dominant systematic uncertainties are the ones related to theory predictions and
MC modelling, while jet uncertainties are the dominant experimental ones.
No significant difference from either the composition or the value of the total uncertainty presented in the published individual analyses is observed.


\section{Exclusion limits}
\label{sec:results}

\begin{figure}[ht!b]
\begin{center}
\begin{subfigure}[t]{0.48\textwidth}
\centering
\includegraphics[width=0.99\textwidth]{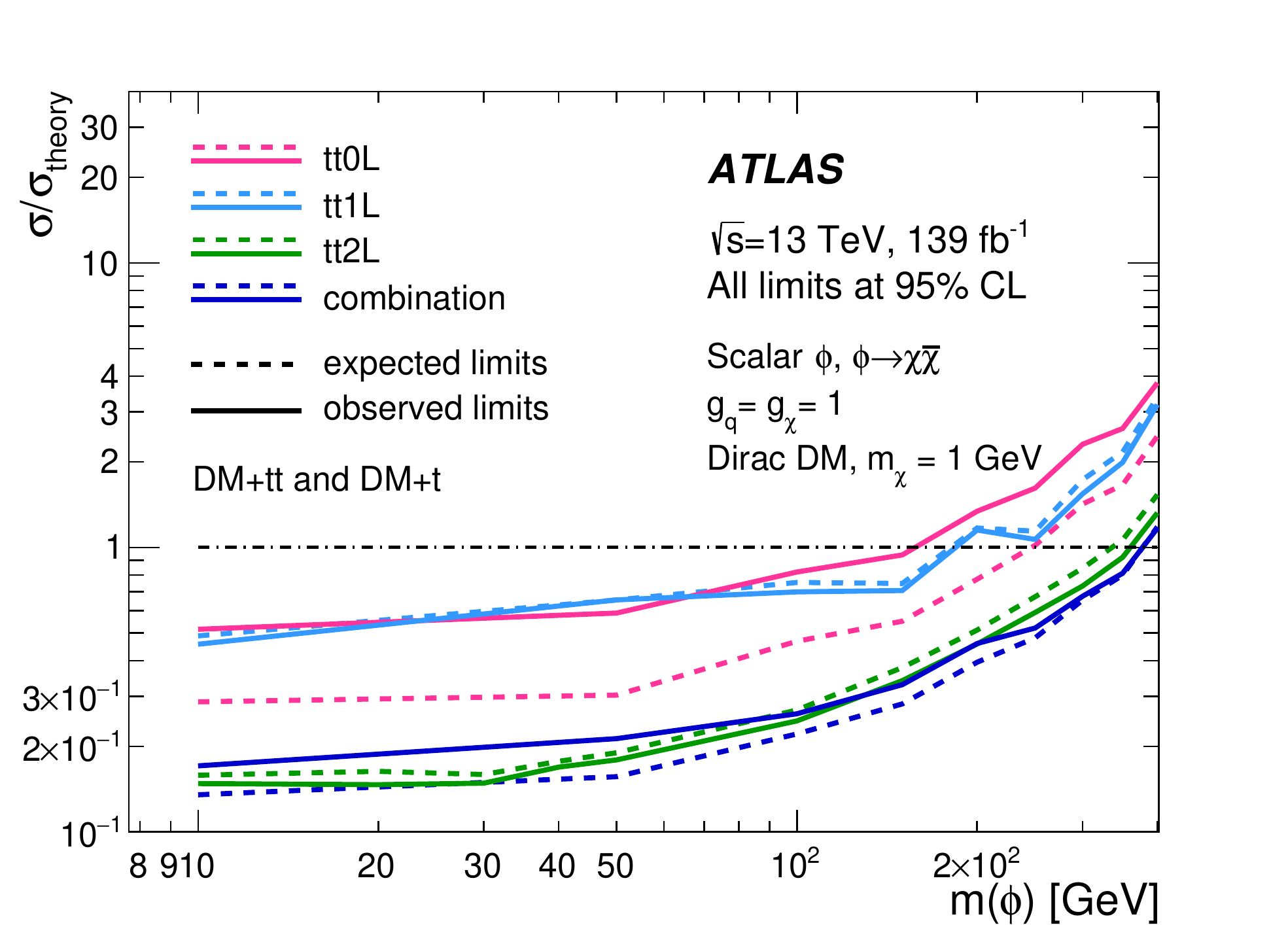}
\caption{}
\label{resscalar}
\end{subfigure}
\begin{subfigure}[t]{0.48\textwidth}
\centering
\includegraphics[width=0.99\textwidth]{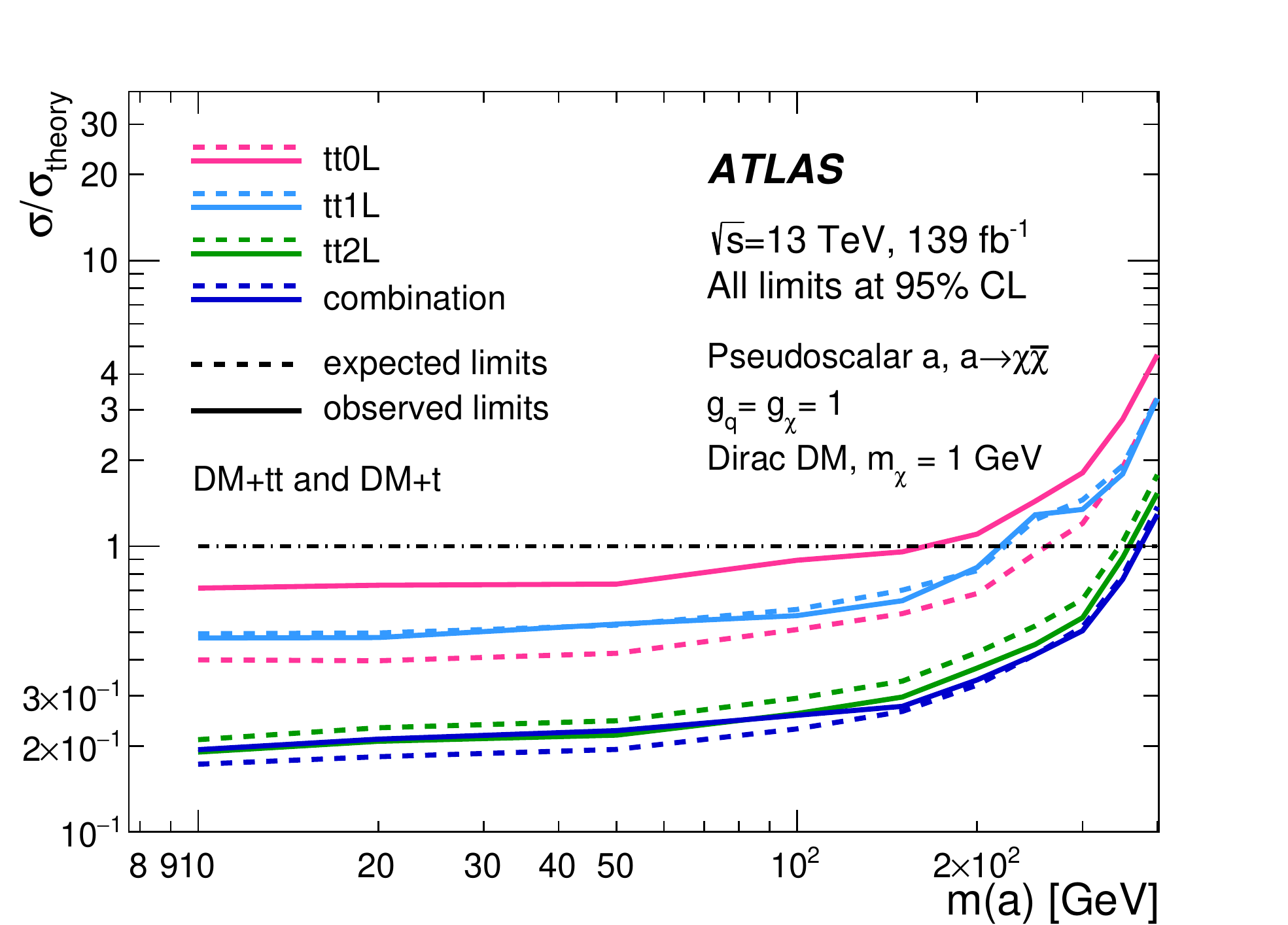}
\caption{}
\label{respseudo}
\end{subfigure}
\caption{Exclusion limits for colour-neutral (a) scalar or (b) pseudoscalar mediator
dark matter models as a function of the mediator mass
$m(\phi)$ or $m(a)$ for a DM mass $m_{\chi} =  1~\GeV$. Associated
production of DM with both single top quarks ($tW$ and $tj$
channels) and top quark pairs is considered. The limits are
calculated at 95\% CL and are expressed in terms of the ratio
of the excluded cross section to the cross section for
a coupling assumption of $g = g_q = g_{\chi} = 1$.  The solid
(dashed) lines show the observed (expected) exclusion limits
for each individual channel and their statistical combination.}
\end{center}
\end{figure}

Exclusion limits at 95\% CL are presented in
Figures~\ref{resscalar}~and~\ref{respseudo} for DM models with a spin-0 scalar or pseudoscalar mediator particle, respectively.
The three individual channels are also presented for comparison.
The \noL\ limits are the result of the statistical combination of
the \noLlow\ and \noLhigh\ SRs.
The \noLlow\ selection improves the expected scalar (pseudoscalar) mediator stand-alone cross-section limit of the \noLhigh\ by
up to 15\% (5\%) and it is strongest for mediator masses values around $10\;\GeV$.
Details of the comparison can be found in Appendix~\ref{sec:lowmetresults}.
 
The signal generation considered in these results includes both the top-quark-pair final states (DM+$\ttbar$) and single-top-quark final
states (DM+$tW$ and DM+$tj$). The limits are expressed in terms of the ratio
of the excluded cross section to the nominal cross section for
a coupling of $g = g_q = g_{\chi} = 1$. With these assumptions, scalar DM models are characterised by
a higher cross section than for pseudoscalar DM models with low mediator masses~\cite{Haisch:2016gry}, while the two
models have very similar cross sections beyond the top quark decay threshold ($m(\phi)$ or $m(a) \sim 2\cdot m_t$).
A DM particle mass of $1\;\GeV$ is considered, although the
results are valid as long as the mass of the mediator is larger than twice the mass of the DM particle. The solid
(dashed) lines show the observed (expected) exclusion limits
for each individual analysis and their statistical
combination. For scalar (pseudoscalar) DM models, the combination extends the
excluded mass range by $50~(25)\;\GeV$ beyond that of the best of the
individual analyses, excluding mediator masses up to $370\;\GeV$.
In addition, the combination improves the expected
cross-section limits by 14\% and 24\%, for low-mass scalar and pseudoscalar
DM mediators, respectively. This directly translates into more stringent
exclusion limits on the couplings.
When only the associated production of DM and two top quarks is considered
in the interpretation of the results, the excluded scalar (pseudoscalar) mediator mass range
obtained from the combination is reduced by
$70\,(20)\,\GeV$ relative to the sensitivity of the combination as reported in Figures~\ref{resscalar}~and~\ref{respseudo}.
As the production of DM in association with a single top quark is most relevant for
higher masses in the scalar mediator models~\cite{Pinna:2017tay},
the impact of this process for masses below $50\;\GeV$ is
negligible. In contrast, for the pseudoscalar mediator models, the ratio of
single-top-quark channel to \ttbar\ channel cross sections is relatively
constant~\cite{Pinna:2017tay}. When considering only DM+$\ttbar$ associated production,
the cross-section upper limit weakens by about 18\% over the
whole mass range.
 
\begin{figure}[ht!b]
\begin{center}
\begin{subfigure}[t]{0.48\textwidth}
\centering
\includegraphics[width=0.99\textwidth]{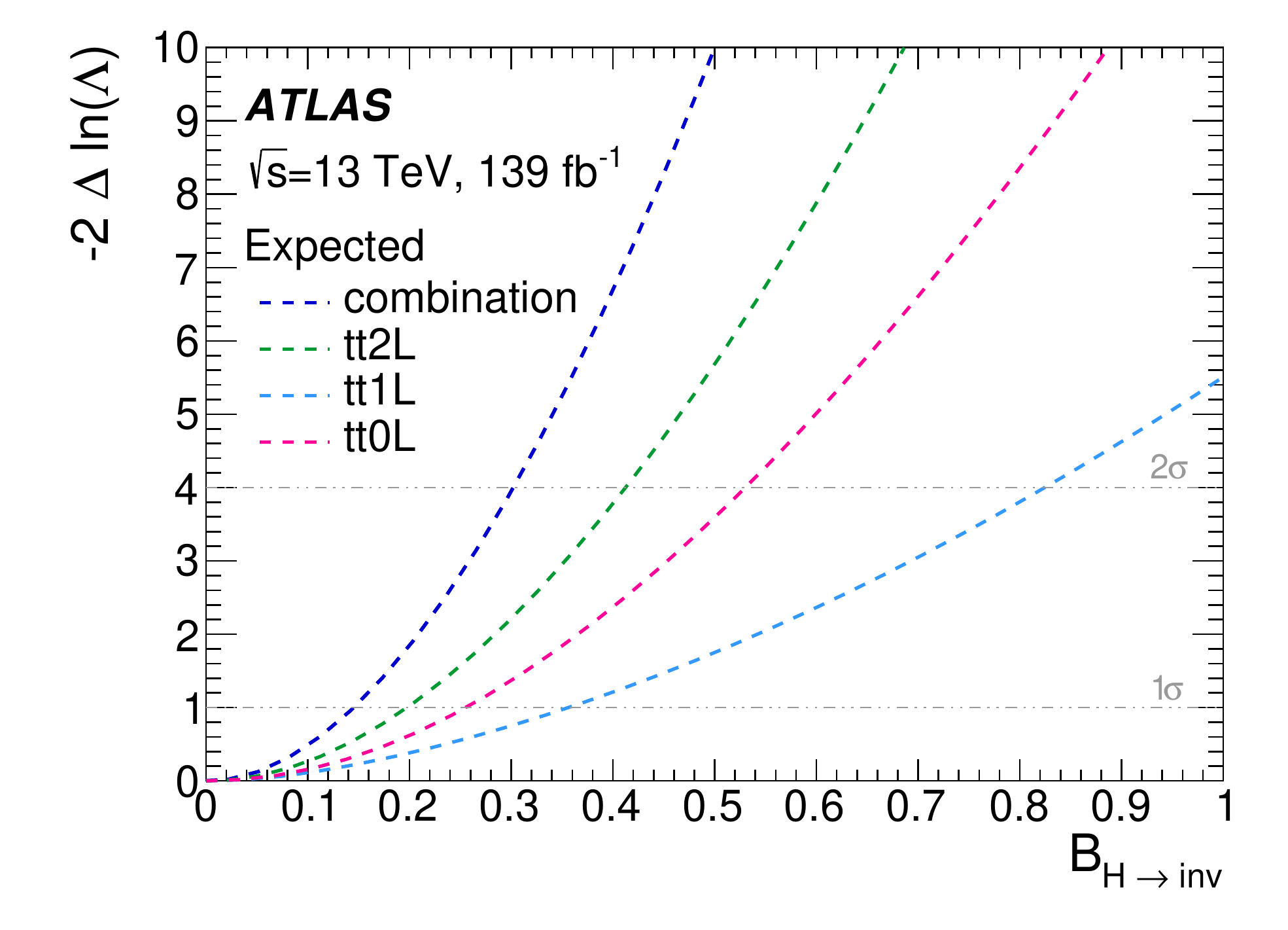}  
\caption{}
\end{subfigure}
\begin{subfigure}[t]{0.48\textwidth}
\centering
\includegraphics[width=0.99\textwidth]{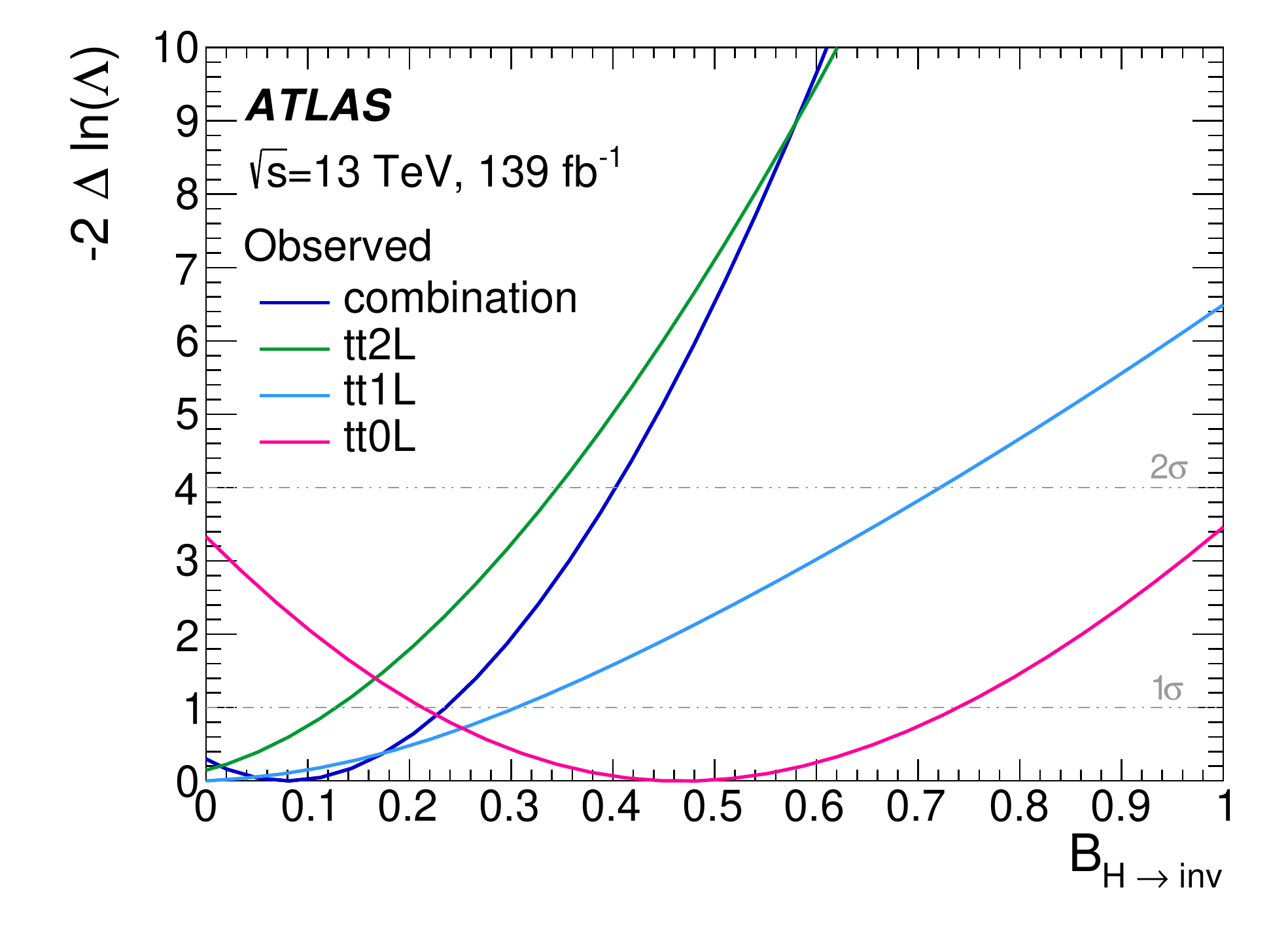}  
\caption{}
\end{subfigure}
\end{center}
\caption{(a) The expected negative logarithmic profile likelihood
ratios $-2\,\Delta\!\ln(\Lambda)$ as a function of \brinv for
each of the three channels and their statistical combination
and (b) these likelihood ratios for the observed data.
\label{fig:lhscan_rtwo}
}
\end{figure}

The negative logarithmic profile likelihood ratios
$-2\,\Delta\!\ln(\Lambda)(\brinv;\theta)$ as a function of \brinv from
the individual analyses and their combination are shown in Figure~\ref{fig:lhscan_rtwo}.
 
Expected results are obtained using the Asimov dataset technique and calculated from asymptotic formulae~\cite{Cowan:2010js}.
The best-fit values of \brinv for the individual analyses are
compatible, agreeing to within one standard deviation. Their statistical combination
yields a best-fit value of $0.08^{+0.15}_{-0.15}$, consistent with the SM prediction of 0.12\%.
The combined observed 95\% CL upper limit on  \brinv  is $0.38$ while
the expected value is $0.30^{+0.13}_{-0.09}$. The individual analysis results are presented in Table~\ref{tab:DMtt_results},
while the details of the \noL\ combination are reported in Appendix~\ref{sec:lowmetresults}.
The overall uncertainty is dominated by the statistical uncertainty of the
data and, to a lesser extent, by systematic uncertainties associated with the modelling
of the SM processes and jet-related uncertainties.
Higgs boson invisible decays represent a specific case of the DM simplified models
considered in the previous section, where the mass of the scalar mediator is assumed
to be $125\;\GeV$. The two results are compatible with each other, when taking into account
the different order of accuracy used in event generation for the \hinv\ model.
 
\begin{table} [htbp!]
\caption{
\label{tab:DMtt_results}
Summary of results from direct searches for invisible decays of the
125~\GeV\ Higgs boson in the \tth topology using 139~\ifb\ of Run~2
data, and their statistical combination. Shown are the best-fit values
of \brinv, computed as not being bounded below by zero, for consistency with previous results~\cite{HIGG-2015-03}.
Observed and expected upper limits on \brinv at the 95\% CL are computed with the $\mathrm{CL_s}$ method and are new results 
with respect to the individual analysis papers quoted in the last table column. The corresponding Asimov datasets for the expected
results are constructed using nuisance parameter values from a fit to
data with $\brinv = 0$, and the quoted uncertainty corresponds to the
68\% confidence interval.
}
\begin{center}
\begin{tabular}{l | c  c  c |l}
\hline
\multirow{2}{*}[1.5mm]{Analysis} & Best fit & Observed & Expected  & \multirow{2}{*}[1.5mm]{Reference} \vspace{-3mm} \\
& \brinv & upper limit & upper limit & \\
\hline
tt0L   		    &      ~~\,$0.48^{+0.27}_{-0.27}$      &     0.95   & $0.52^{+0.23}_{-0.16}$  & \cite{SUSY-2018-12}, this document \\
tt1L   		    &   $-0.04^{+0.35}_{-0.29}$       &     0.74   & $0.80^{+0.40}_{-0.26}$  & \cite{SUSY-2018-07}, this document \\
tt2L    	            &    $-0.08^{+0.20}_{-0.19}$    &     0.36   & $0.40^{+0.18}_{-0.12}$  & \cite{SUSY-2018-08}, this document \\
\hline
\(\ttbar H\) comb. &       ~~\,$0.08^{+0.15}_{-0.15}$     &    0.38   & $0.30^{+0.13}_{-0.09}$ & This document \\
\hline
\end{tabular}
\end{center}
\end{table}


\FloatBarrier
 
\section{Conclusion}
\label{sec:conclusion}
 
In summary, a statistical combination of three analyses using 139~fb$^{-1}$ of $pp$ collisions delivered by the LHC
at a centre-of-mass energy of 13\;\TeV\ and collected by the ATLAS detector is presented.
The three analyses are all designed to select events with two top quarks and invisible particles, and consider all
possible light lepton multiplicities arising from the decays of the two top quarks.
 
The statistical combination is used
to set 95\% confidence-level constraints on spin-0 simplified
dark matter models. All production modes with top quarks  in the final state (DM+$\ttbar$, DM+$t$) are considered.
For scalar (pseudoscalar) dark matter models, the combination extends the excluded mass range by
$50~(25)\;\GeV$ beyond that of the best of the individual channels, excluding mediator masses up to $370\;\GeV$ with all couplings set to unity. In addition, the combination improves the observed coupling exclusion limit by 24\%, assuming a pseudoscalar mediator mass of $10\;\GeV$.
 
The specific case where the mediator corresponds to the SM $125~\GeV$ Higgs boson
is also considered when interpreting the results presented in this paper.
An upper limit on the Higgs boson invisible branching ratio of 0.38 ($0.30^{+0.13}_{-0.09}$) is observed (expected) at 95\% confidence level.

\section*{Acknowledgements}


We thank CERN for the very successful operation of the LHC, as well as the
support staff from our institutions without whom ATLAS could not be
operated efficiently.
 
We acknowledge the support of
ANPCyT, Argentina;
YerPhI, Armenia;
ARC, Australia;
BMWFW and FWF, Austria;
ANAS, Azerbaijan;
CNPq and FAPESP, Brazil;
NSERC, NRC and CFI, Canada;
CERN;
ANID, Chile;
CAS, MOST and NSFC, China;
Minciencias, Colombia;
MEYS CR, Czech Republic;
DNRF and DNSRC, Denmark;
IN2P3-CNRS and CEA-DRF/IRFU, France;
SRNSFG, Georgia;
BMBF, HGF and MPG, Germany;
GSRI, Greece;
RGC and Hong Kong SAR, China;
ISF and Benoziyo Center, Israel;
INFN, Italy;
MEXT and JSPS, Japan;
CNRST, Morocco;
NWO, Netherlands;
RCN, Norway;
MEiN, Poland;
FCT, Portugal;
MNE/IFA, Romania;
MESTD, Serbia;
MSSR, Slovakia;
ARRS and MIZ\v{S}, Slovenia;
DSI/NRF, South Africa;
MICINN, Spain;
SRC and Wallenberg Foundation, Sweden;
SERI, SNSF and Cantons of Bern and Geneva, Switzerland;
MOST, Taiwan;
TENMAK, T\"urkiye;
STFC, United Kingdom;
DOE and NSF, United States of America.
In addition, individual groups and members have received support from
BCKDF, CANARIE, Compute Canada and CRC, Canada;
PRIMUS 21/SCI/017 and UNCE SCI/013, Czech Republic;
COST, ERC, ERDF, Horizon 2020 and Marie Sk{\l}odowska-Curie Actions, European Union;
Investissements d'Avenir Labex, Investissements d'Avenir Idex and ANR, France;
DFG and AvH Foundation, Germany;
Herakleitos, Thales and Aristeia programmes co-financed by EU-ESF and the Greek NSRF, Greece;
BSF-NSF and MINERVA, Israel;
Norwegian Financial Mechanism 2014-2021, Norway;
NCN and NAWA, Poland;
La Caixa Banking Foundation, CERCA Programme Generalitat de Catalunya and PROMETEO and GenT Programmes Generalitat Valenciana, Spain;
G\"{o}ran Gustafssons Stiftelse, Sweden;
The Royal Society and Leverhulme Trust, United Kingdom.
 
The crucial computing support from all WLCG partners is acknowledged gratefully, in particular from CERN, the ATLAS Tier-1 facilities at TRIUMF (Canada), NDGF (Denmark, Norway, Sweden), CC-IN2P3 (France), KIT/GridKA (Germany), INFN-CNAF (Italy), NL-T1 (Netherlands), PIC (Spain), ASGC (Taiwan), RAL (UK) and BNL (USA), the Tier-2 facilities worldwide and large non-WLCG resource providers. Major contributors of computing resources are listed in Ref.~\cite{ATL-SOFT-PUB-2021-003}.


\clearpage
\appendix
\part*{Appendix}
\addcontentsline{toc}{part}{Appendix}
 
\section{The \noLlow\ analysis}
\label{sec:lowmetapp}


The \noLlow\ analysis aims to enhance the sensitivity to DM+\ttbar signals with low mediator masses ($m(\phi),m(a)<100~\GeV$).
Two main discriminating variables, \coshmax\ and \chitwotop, are defined in order to reduce the most dominant top quark backgrounds.
Angular separations between $b$-tagged jets, \met or large-radius jets are used to further reduce the contamination from Standard Model processes.
To ensure orthogonality with the \noLhigh\ selections, additional orthogonalisation requirements are also applied, as detailed in Section~\ref{sec:analyses_detail}.
 
\subsection{Discriminating variables}
\label{app:var}
 
The full event selections performed in the signal regions can be found in Table~\ref{tab:lowmet_sr}. The discriminating variables are described in more detail below.
 
\subsubsection*{\coshmax}

The \coshmax\ variable is designed to distinguish signal events from single-top events in the $tW$ channel and \ttbar\ events with a lepton missed by the reconstruction algorithms (top-with-lost-lepton), which are among the main backgrounds in the analysis. Such events may enter the signal regions because of high \met originating from the $t \to bW \to b\ell\nu$ decay, and the lost lepton.
 
The reconstruction of events containing a top quark with a lost lepton is attempted by assuming that the \met is equal to the \pt of the leptonically decaying $W$ boson with a lost lepton, $E_{\text{T}}^{\text{miss}} \sim p_{\text{T}}^{W}$.
 
The top-with-lost-lepton background can then be reconstructed by combining the missing transverse momentum with the correct $b$-tagged jet ($t\to bW$).
In practice, a four-vector with \pt and $\phi$ corresponding to the $\vec{p}_{\text{T}}^{\text{miss}}$ vector and its mass equal to the $W$ boson mass is built, while
its pseudorapidity $\eta_W$ (or equivalently $p_{z}^{W}$) remains unknown. Choosing the $x$-axis to be in the direction of $\pt^{W}$ and adopting $(E, p_x, p_y, p_z)$ coordinates:
 
\begin{align}
\vec{p}_{W} &= \left(\sqrt{(\pt^{W})^2+(p_{z}^{W})^2+m_{W}^2},\pt^{W}, 0, p_{z}^{W}\right), \label{eq:pW} \\
\vec{p}_{b} &= \left(\sqrt{(\pt^{b})^2+(p_{z}^{b})^2+m_{b}^2},\pt^{b}\cdot\cos(\phi_{W}-\phi_{b}), \pt^{b}\cdot\sin(\phi_{W}-\phi_{b}), p_{z}^{b}\right), \label{eq:pb} \\
m_{t}^2 &= (\vec{p}_{W} + \vec{p}_{b})^2, \label{eq:mt}
\end{align}
 
where the $b$ superscript and subscript refer to one of the selected $b$-tagged jets.
Substituting Eqs.~(\ref{eq:pW}) and~(\ref{eq:pb}) in Eq.~(\ref{eq:mt}),
and assuming the massless limit for the $b$-tagged jet, the equivalence below is formed:
 
\begin{align}
\sqrt{1+\left(\frac{m_{W}}{\pt^{W}\cdot\cosh\eta_{W}}\right)^2}\cdot\cosh\eta_{W}\cdot\cosh\eta_{b}-\sinh\eta_{W}\cdot\sinh\eta_{b}
= \frac{m_{t}^2-m_{W}^2}{2\pt^{W}\pt^{b}}+\cos(\phi_{W}-\phi_{b}),
\label{eq:coshmax_long}
\end{align}
 
where $\eta_{W}$ is unknown. Given that $\met \sim \pt^{W} > 160~\GeV$ in the signal regions and $\cosh\eta_{W}\ge1$, it may be assumed that
$m_{W} \sim 80~\GeV \ll \pt^{W}\cdot\cosh\eta_{W}$, such that:
 
\begin{align}
\sqrt{1+\left(\frac{m_{W}}{\pt^{W}\cdot\cosh\eta_{W}}\right)^2} \sim 1. \nonumber
\end{align}
 
Equation~(\ref{eq:coshmax_long}) can thus be simplified:
 
\begin{align}
\cosh(\eta_{W}-\eta_{b}) &\sim \frac{m_{t}^2-m_{W}^2}{2\pt^{W}\pt^{b}}+\cos(\phi_{W}-\phi_{b}) \label{eq:cosh} \\ \nonumber
&\sim \frac{m_{t}^2-m_{W}^2}{2\met\pt^{b}}+\cos(\phi_{\met}-\phi_{b}).
\end{align}
 
By definition, $\cosh(x) \geq 1$ so that the right-hand side of Eq.~(\ref{eq:cosh}) is expected to be larger than 1 in the case of successful leptonic top reconstruction. The discriminating observable \coshmax is therefore defined as:
 
\begin{align}
\coshmax = \max\{\cosh\Delta\eta^{1}_{W,b}, \cosh\Delta\eta^{2}_{W,b}\}, \nonumber
\end{align}
 
where $\Delta\eta^{1}_{W,b}$ and $\Delta\eta^{2}_{W,b}$ represent the pseudorapidity difference between the $W$ boson candidate and either of the two leading $b$-tagged jets selected in the event. Events with high \coshmax\ values are likely to contain a top quark with a lost lepton and are excluded from the signal regions.
 
\begin{figure}[ht!b]
\begin{center}
\begin{subfigure}[t]{0.5\textwidth}
\includegraphics[width=0.99\textwidth]{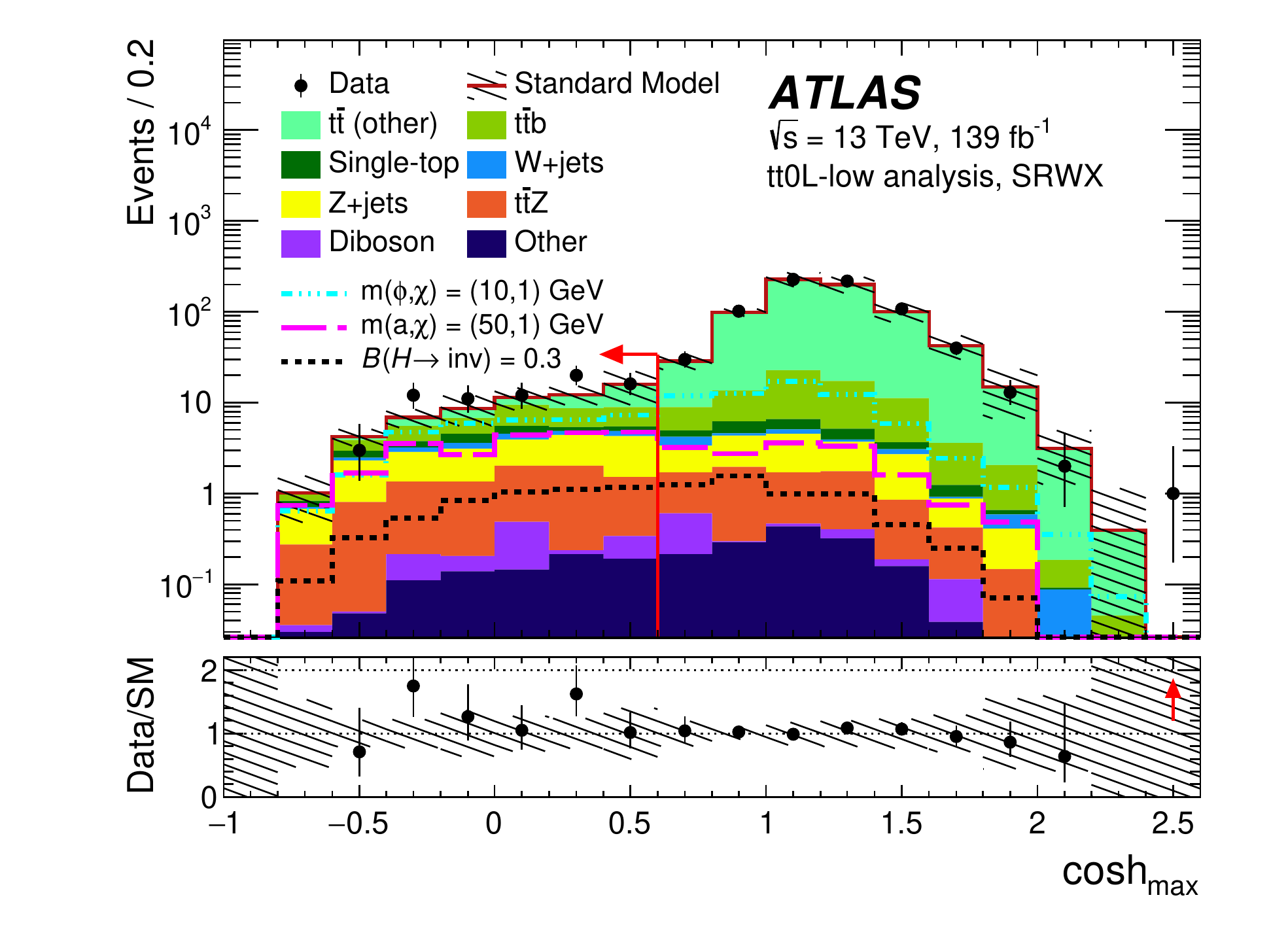}
\caption{}
\end{subfigure}
\begin{subfigure}[t]{0.5\textwidth}
\includegraphics[width=0.99\textwidth]{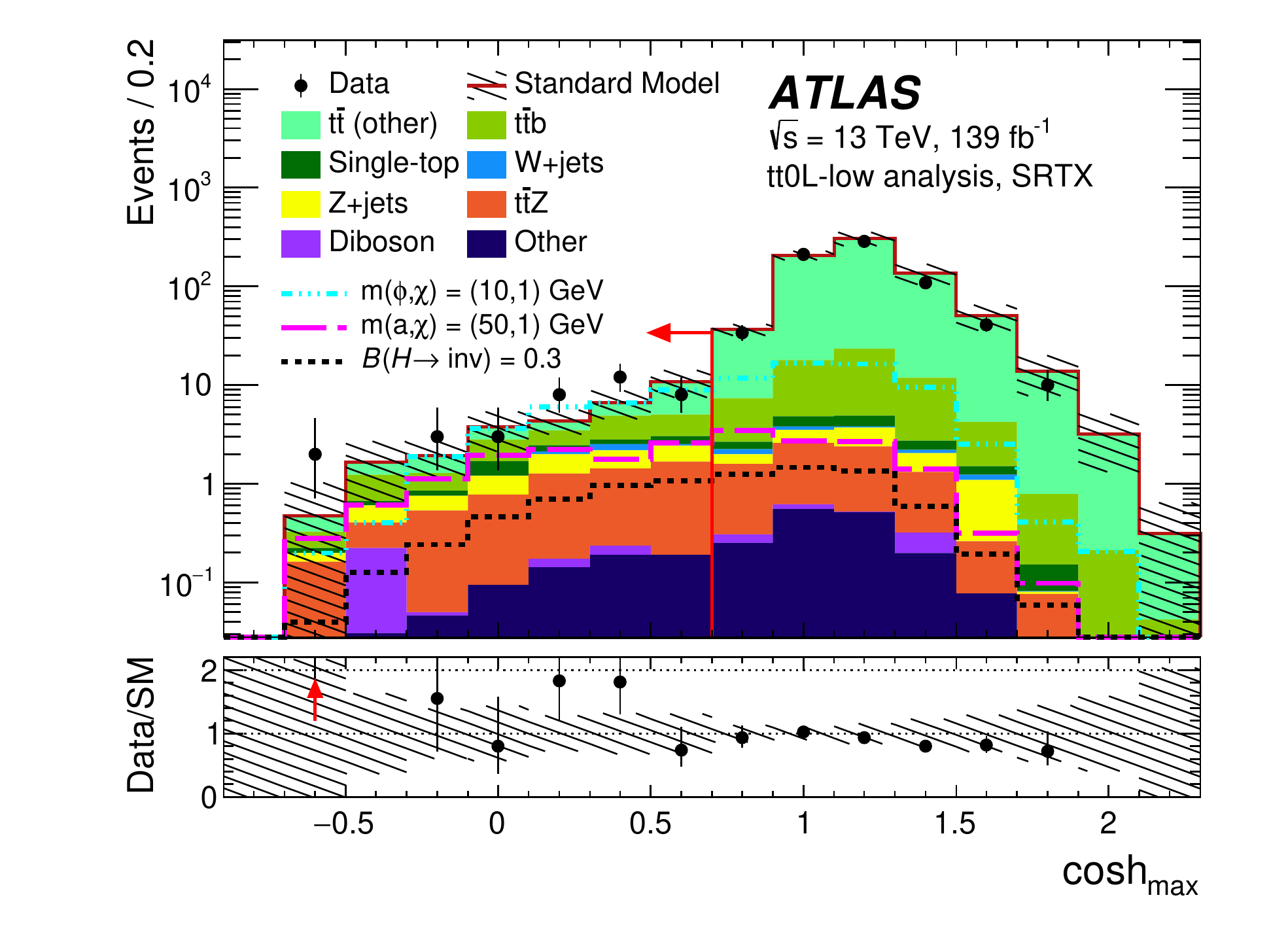}
\caption{}
\end{subfigure}
\caption{Distributions of \coshmax in (a) SRWX and (b) SRTX events passing all the SR requirements
except those on \coshmax itself (which are indicated by the arrows).
The contributions from all SM backgrounds are shown after the profile likelihood simultaneous fit to all \noLlow\ CRs,
with the hatched bands representing the total uncertainty.
The category `\ttbar\ (other)' represents \ttbar\ events without extra jets or events with extra light-flavour jets.
`Other' includes contributions from $\ttbar{+}W$, $tZ$ and $tWZ$ processes.
The expected distributions for selected signal models are shown as dashed lines.
The underflow (overflow) events are included in the first (last) bin.
The bottom panels show the ratio of the observed data to the total SM background prediction,
with the hatched area representing the total uncertainty in the background prediction
and the red arrows marking data outside the vertical-axis range.}
\label{fig:cosh_app}
\end{center}
\end{figure}

Figure~\ref{fig:cosh_app} illustrates the modelling of the shape of \coshmax\ in SRWX and SRTX.
The \coshmax\ distribution in SR0X is shown in Figure~\ref{fig:key_discriminants}.
 
\subsubsection*{\chitwotop}
 
The \chitwotop\ observable approximately quantifies how likely an event is to include two hadronically decaying top quarks.
It is used primarily to reject backgrounds containing no hadronic top quark decays, such as $Z$+jets events. It is defined as follows:
 
\begin{align}
\chitwotop &= \left(\frac{m_{W_1}-m_{W_{\text{ref}}}}{\sigma_{m_{W}}}\right)^2 \label{eq:chi2} \\ \nonumber
&\quad + \left(\frac{(m_{t_1}-m_{W_1})-(m_{t_{\text{ref}}}-m_{W_{\text{ref}}})}{\sigma_{m_{t}-m_{W}}}\right)^2 \\ \nonumber
&\quad + \left(\frac{(m_{t_2}-m_{W_2})-(m_{t_{\text{ref}}}-m_{W_{\text{ref}}})}{\sigma_{m_{t}-m_{W}}}\right)^2.
\end{align}
 
Up to seven jets, including the two selected $b$-tagged jets, are considered in the calculation. The first $W$ boson candidate, $W_1$, is built from two non-$b$-tagged jets,
while the first top quark candidate, $t_1$, combines $W_1$ and one of the $b$-tagged jets, $b_1$, such that $t_1\to W_1 b_1$. According to Monte Carlo simulations, the second $W$ boson candidate, $W_2$, is in more than 50\% of the cases too soft to lead to two individual jets satisfying the reconstruction criteria. Hence, it is built from a single non-$b$-tagged jet to which the mass of the $W$ boson is attributed. As a result, the second top quark candidate, $t_2 \to W_2 b_2$, contains only one non-$b$-tagged jet and the remaining $b$-tagged jet, $b_2$.
 
The first term in Eq.~(\ref{eq:chi2}) corresponds to the invariant mass constraint from $W_1$. The values $m_{W_{\text{ref}}}$ and $\sigma_{m_{W}}$ are respectively the mean and the standard deviation of
the experimental invariant mass distribution expected for hadronically decaying $W$ bosons. The second and third terms correspond to the invariant mass constraints from $t_1$ and $t_2$, respectively. Since $m_{W_1}$ and $m_{t_1}$ ($m_{W_2}$ and $m_{t_2}$) are
strongly correlated, the $W$ boson mass is subtracted from the top quark mass to decouple these two terms from the first one. The values of $m_{t_{\text{ref}}}$ and $\sigma_{m_{t}-m_{W}}$ are respectively the mean of the experimental top quark mass distribution and the standard deviation of the  $m_{t}-m_{W}$ distribution expected for reconstructed hadronic top quark decays. The values of $m_{W_{\text{ref}}}$, $\sigma_{m_{W}}$, $m_{t_{\text{ref}}}$ and $\sigma_{m_{t}-m_{W}}$ are taken from Ref.~\cite{ATLAS-CONF-2018-045}:
 
\begin{itemize}
\item{} $m_{W_{\text{ref}}} = 80.51~\GeV$, $\sigma_{m_{W}} = 12.07~\GeV$,
\item{} $m_{t_{\text{ref}}}-m_{W_{\text{ref}}} = 85.17~\GeV$, $\sigma_{m_{t}-m_{W}} = 16.05~\GeV$.
\end{itemize}
 
The $\chi^2$ is recomputed for each possible jet combination and the final \chitwotop\ corresponds to the minimum value obtained. Events with high \chitwotop\ values are less likely to contain two hadronic top quark decays and are therefore excluded from the signal regions.
 
\begin{figure}[ht!b]
\begin{center}
\begin{subfigure}[t]{0.5\textwidth}
\includegraphics[width=0.99\textwidth]{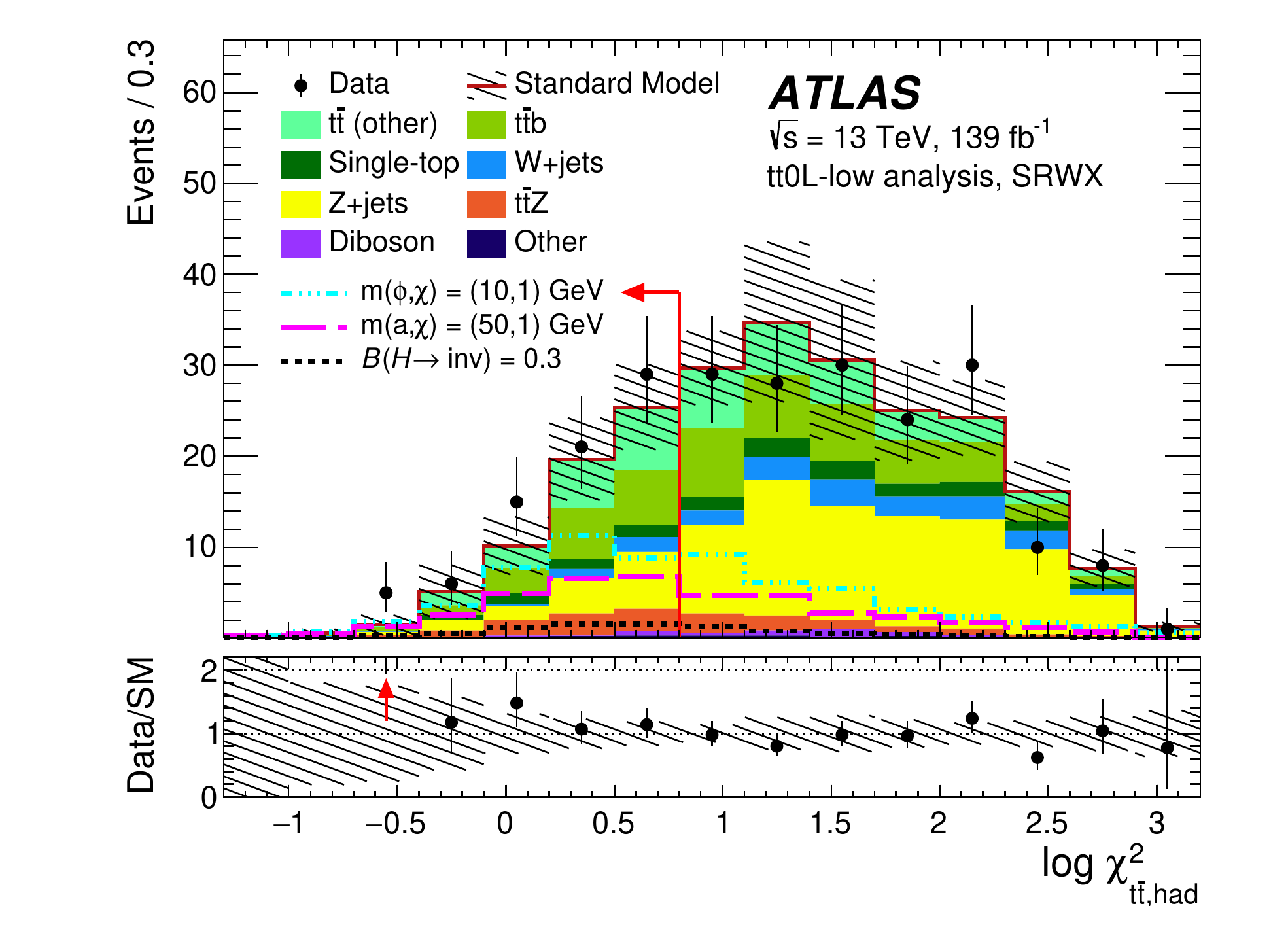}
\caption{}
\end{subfigure}
\begin{subfigure}[t]{0.5\textwidth}
\includegraphics[width=0.99\textwidth]{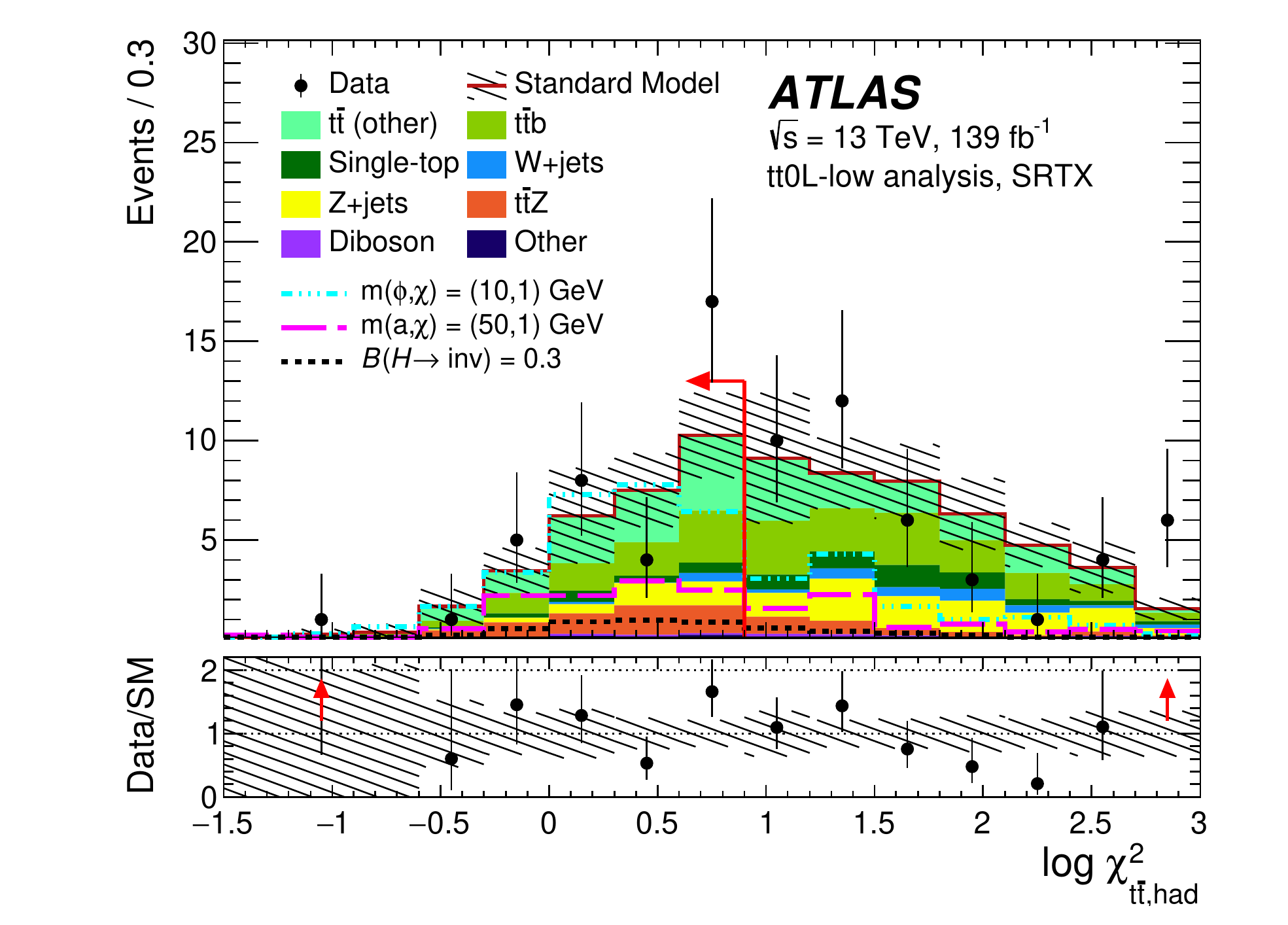}
\caption{}
\end{subfigure}
\caption{Distributions of \chitwotop in (a) SRWX and (b) SRTX events passing all the SR requirements
except those on \chitwotop itself (which are indicated by the arrows).
The contributions from all SM backgrounds are shown after the profile likelihood simultaneous fit to all \noLlow\ CRs,
with the hatched bands representing the total uncertainty.
The category `\ttbar\ (other)' represents \ttbar\ events without extra jets or events with extra light-flavour jets.
`Other' includes contributions from $\ttbar{+}W$, $tZ$ and $tWZ$ processes.
The expected distributions for selected signal models are shown as dashed lines.
The underflow (overflow) events are included in the first (last) bin.
The bottom panels show the ratio of the observed data to the total SM background prediction,
with the hatched area representing the total uncertainty in the background prediction
and the red arrows marking data outside the vertical-axis range.}
\label{fig:chi2_app}
\end{center}
\end{figure}

Figure~\ref{fig:chi2_app} illustrates the modelling of the shape of \chitwotop\ in SRWX and SRTX.
The \chitwotop\ distribution in SR0X is shown in Figure~\ref{fig:key_discriminants}.
 
\subsection{Background estimation}
\label{app:bkg}
 
The event topologies in the signal regions and control regions are kept as similar as possible to reduce any bias originating from differences between their kinematic phase spaces. For this purpose, control regions with one or more leptons in the final state are split according to the mass of the heaviest large-radius jet, as is done for the signal regions, while all \met-related variables are recalculated by treating the selected leptons as invisible, denoted by the subscript `no lepton' in the variable names.
 
One of the most prominent sources of background in the signal regions is semileptonic \ttbar\ decay
where the lepton is misreconstructed or outside the detector acceptance, while the contribution from the dileptonic \ttbar\ decay is negligible.
Control regions selecting events with exactly one lepton ($e$ or $\mu$) are defined
in order to estimate the background originating from a top quark decay with a lost lepton,
which includes single-top events in the $tW$ channel, and $\ttbar{+}b$ and \ttbar\ (other) events.
 
A $\chi^2$-based observable~\cite{EXOT-2015-04}, \chileptop, taking into account the kinematic properties of \met, lepton, jets and the $b$-tagging information, is used to reconstruct semileptonic \ttbar events and separate them from $tW$ and $\ttbar{+}b$ events. It follows an approach similar to that for the \chitwotop\ variables by placing constraints on the masses of the hadronically decaying $W$ boson, the hadronically decaying top quark and the leptonically decaying top quark. The presence of extra $b$-tagged jets is used to select $\ttbar{+}b$ over single-top processes. Tighter \coshmaxnolepton selections are required in the single-top control regions to reduce the contamination from semileptonic \ttbar events failing the \chileptop reconstruction and attain high purity in $tW$ events. Table~\ref{tab:lowmet_1lcr}
presents the full event selections applied to define the top-with-lost-lepton control regions.

\begin{table}[!htb]
\centering
\scriptsize
\caption{Selection criteria for the top-with-lost-lepton control regions used in the \noLlow\ analysis.}
\label{tab:lowmet_1lcr}
\begin{tabular}{|>{\centering}p{0.24\textwidth-2\tabcolsep}|>{\centering}p{0.3\textwidth-2\tabcolsep}|>{\centering}p{0.15\textwidth-2\tabcolsep}|>{\centering}p{0.15\textwidth-2\tabcolsep}|>{\centering\arraybackslash}p{0.15\textwidth-2\tabcolsep}|}
\hline
\multirow{13}{*}{Shared selections} & \textbf{Variables} & \textbf{CR0X} & \textbf{CRWX} & \textbf{CRTX} \\
\cline{2-5}
& $N_{\textrm{lepton}}$ & \multicolumn{3}{c|}{$=$ 1} \\
\cline{2-5}
& \metnolepton[\GeV] & \multicolumn{3}{c|}{$>$ 160} \\
\cline{2-5}
& \met[\GeV] & \multicolumn{3}{c|}{$<$ 250, when passing $b$-jet triggers} \\
\cline{2-5}
& $\mathcal{S}_{\textrm{no lepton}}$ & \multicolumn{3}{c|}{$>$ 10} \\
\cline{2-5}
& $\Delta\phi_{\min}(\vec{p}_{\text{T,1-4}},\vec{p}_{\text{T,no lepton}}^{\text{miss}})$ & $>$ 1.0 & \multicolumn{2}{c|}{$>$ 0.5} \\
\cline{2-5}
& \drbb & \multicolumn{3}{c|}{$>$ 1.2} \\
\cline{2-5}
& $N_{\textrm{large-radius jet}}$ & $=$ 0 & \multicolumn{2}{c|}{$>$ 0} \\
\cline{2-5}
& $m_{\textrm{large-radius jet}}$ [\GeV] & --- & (40, 130) & $\geq$ 130 \\
\cline{2-5}
& $\Delta R_{\min}$(large-radius jet, $b$-tagged jets) & \multicolumn{2}{c|}{---} & $<$ 1.2 \\
\cline{2-5}
& \coshmaxnolepton & $<$ 0.9 & $<$ 0.95 & $<$ 1.0 \\
\cline{2-5}
& \chitwotop & $<$ 10 & $<$ 20 & $<$ 40 \\
\cline{2-5}
& $\pt^{\ttbar}/\metnolepton$ & (0.7, 1.2) & \multicolumn{2}{c|}{(0.5, 1.2)} \\
\hline
\hline
\multirow{2}{*}{\ttbar\ (other) enriched selections} & \textbf{Variables} & \textbf{CR0X}$_{\ttbar}$ & \textbf{CRWX}$_{\ttbar}$ & \textbf{CRTX}$_{\ttbar}$ \\
\cline{2-5}
& \chileptop & \multicolumn{3}{c|}{$<$ 6} \\
\hline
\hline
\multirow{3}{*}{$\ttbar{+}b$ enriched selections} & \textbf{Variables} & \textbf{CR0X}$_{\ttbar{+}b}$ & \textbf{CRWX}$_{\ttbar{+}b}$ & \textbf{CRTX}$_{\ttbar{+}b}$ \\
\cline{2-5}
& \chileptop & \multicolumn{3}{c|}{$\geq$ 6} \\
\cline{2-5}
& $N_{\textrm{extra }b\textrm{-tagged jet}}$ & \multicolumn{3}{c|}{$\geq$ 1} \\
\hline
\hline
\multirow{4}{*}{Single-top enriched selections} & \textbf{Variables} & \textbf{CR0X}$_{\textrm{single-top}}$ & \textbf{CRWX}$_{\textrm{single-top}}$ & \textbf{CRTX}$_{\textrm{single-top}}$ \\
\cline{2-5}
& \chileptop & \multicolumn{3}{c|}{$\geq$ 30} \\
\cline{2-5}
& $N_{\textrm{extra }b\textrm{-tagged jet}}$ & \multicolumn{3}{c|}{$=$ 0} \\
\cline{2-5}
& \coshmaxnolepton & $<$ 0.5 & $<$ 0.6 & $<$ 0.7 \\
\hline
\end{tabular}
\end{table}
 
Another major background component in the signal regions contains $Z\to\nu\nu$ produced in association with jets.
Control regions selecting events with two leptons with opposite charge and the same flavour ($ee$ or $\mu\mu$)
are defined in order to estimate the $Z(\nu\nu)$+jets background.
The invariant mass and transverse momentum of the dilepton system, $m_{\ell\ell}$ and $\pt^{\ell\ell}$ respectively, and the missing transverse momentum significance $\mathcal{S}$ serve as the major discriminants
to suppress the contamination from dileptonic \ttbar\ events.
To obtain enough events, several selections applied in the signal regions are omitted in the corresponding CRs.
Table~\ref{tab:lowmet_2lcr} presents the full event selections applied to define the $Z$+jets control regions.
 
\begin{table}[!htb]
\centering
\scriptsize
\caption{Selection criteria for the $Z$+jets control regions used in the \noLlow\ analysis.}
\label{tab:lowmet_2lcr}
\begin{tabular}{|>{\centering}p{0.3\textwidth-2\tabcolsep}|>{\centering}p{0.23\textwidth-2\tabcolsep}|>{\centering}p{0.23\textwidth-2\tabcolsep}|>{\centering\arraybackslash}p{0.23\textwidth-2\tabcolsep}|}
\hline
\textbf{Variables} & \textbf{CR0X}$_{\Zjets}$ & \textbf{CRWX}$_{\Zjets}$ & \textbf{CRTX}$_{\Zjets}$ \\
\hline
$N_{\textrm{lepton}}$ & \multicolumn{3}{c|}{$=$ 2} \\
\hline
Orthogonalisation & \multicolumn{3}{c|}{N$_{\textrm{large-radius jet}}^{R=1.2} < 2$ \hspace{0.01\textwidth} or \hspace{0.01\textwidth} $m_{\textrm{subleading large-radius jet}}^{R=1.2} < 60~\GeV$} \\
\hline
\metnolepton[\GeV] & \multicolumn{3}{c|}{$>$ 160} \\
\hline
$\mathcal{S}_{\textrm{no lepton}}$ & \multicolumn{3}{c|}{$>$ 8} \\
\hline
$\Delta\phi_{\min}(\vec{p}_{\text{T,1-4}},\vec{p}_{\text{T}}^{\text{miss}})$ & \multicolumn{3}{c|}{$>$ 0.5} \\
\hline
$N_{\textrm{large-radius jet}}$ & $=$ 0 & \multicolumn{2}{c|}{$>$ 0} \\
\hline
$m_{\textrm{large-radius jet}}$ [\GeV] & --- & (40, 130) & $\geq$ 130 \\
\hline
\hline
$m_{\ell\ell}$ [\GeV] & \multicolumn{3}{c|}{(80, 100)} \\
\hline
$\pt^{\ell\ell}$ [\GeV] & \multicolumn{3}{c|}{$>$ 160} \\
\hline
$\mathcal{S}$ & \multicolumn{3}{c|}{$<$ 5} \\
\hline
\end{tabular}
\end{table}
 
Validation regions are not included in the statistical model and serve only to validate the extrapolation over lepton multiplicity when going from the control regions to the signal regions. The event selections for the validation regions therefore require zero leptons, while being orthogonal to the signal region selections.
 
In the \ttbar-enriched validation regions, \ttbar events are selected by inverting
the tight \coshmax requirement applied in the signal regions and adding a looser upper bound.
The validation regions for $\ttbar{+}b$, single-top and $Z$+jets are merged into a
single \ttbar-suppressed validation region because of the limited number of events in the 0-lepton phase space.
In these regions the \chitwotop selection applied in the signal regions is inverted.
The $\pt^{\ttbar}/\met$ requirements are discarded because they become irrelevant when the value of \chitwotop is too large.
Tight \drbb selections are imposed to minimise the contamination from \Wjets events,
with their thresholds optimised in each region to provide a number of events similar to that in the \ttbar-enriched VRs.
All the background predictions in the VRs agree with the data to within $1\sigma$.

\subsection{Results}
\label{sec:lowmetresults}

All \noLlow\ signal and control regions are included in a statistical model based on the combined likelihood fit. The normalisations of the $\ttbar{+}b$, \ttbar\ (other), single-top and $Z$+jets background processes are free-floating. For the \ttbar\ background, the normalisation factors are decorrelated in the three kinematic regimes (CR0X, CRWX and CRTX) to account for a possible top quark \pt\ dependence of the normalisation factor. The yield results are presented in Table~\ref{tab:lowmetyield}.
 
\begin{figure}[!htbp]
\begin{center}
\begin{subfigure}[t]{0.5\textwidth}
\includegraphics[width=0.99\textwidth]{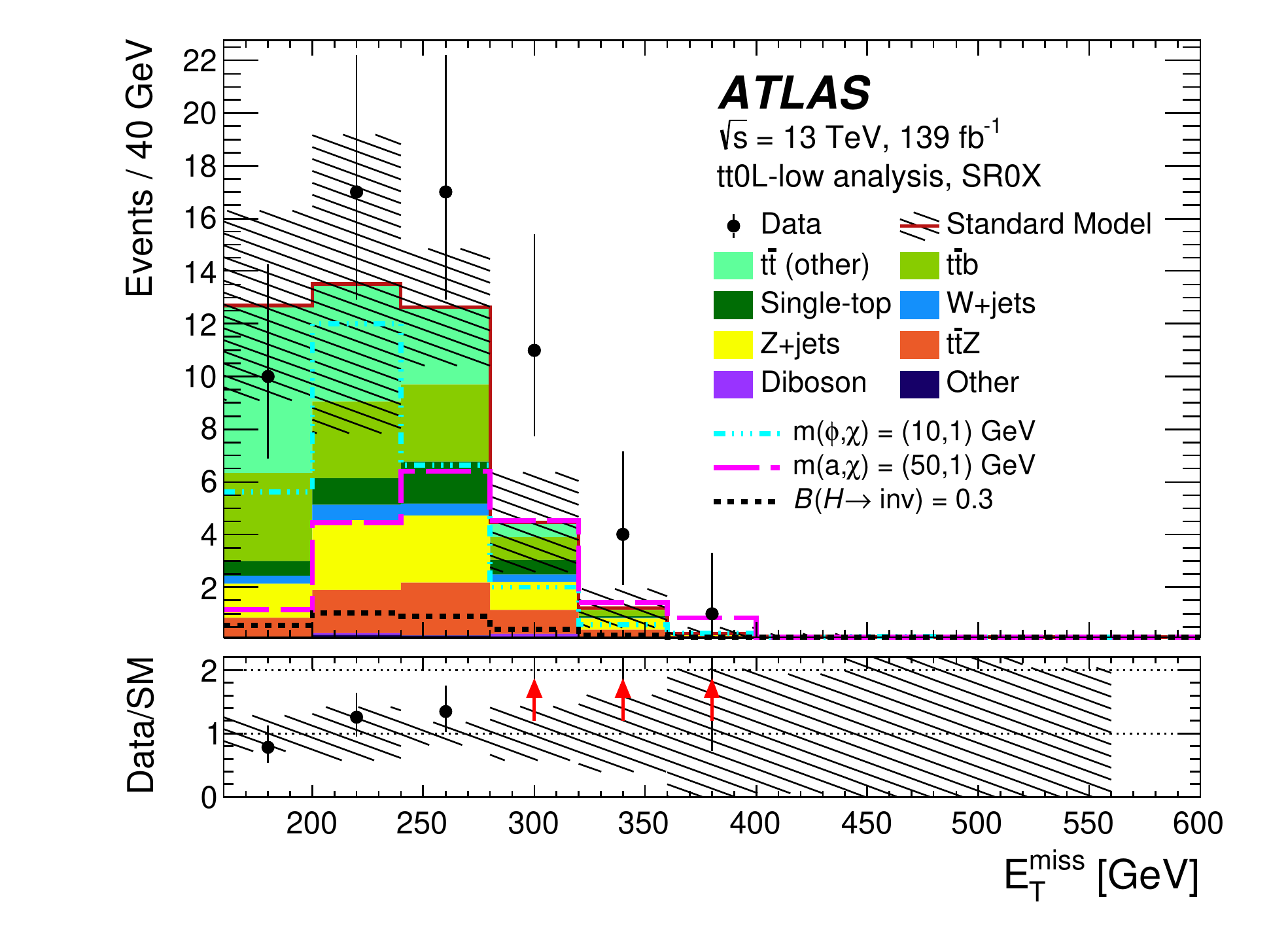}
\caption{}
\end{subfigure}
\begin{subfigure}[t]{0.5\textwidth}
\includegraphics[width=0.99\textwidth]{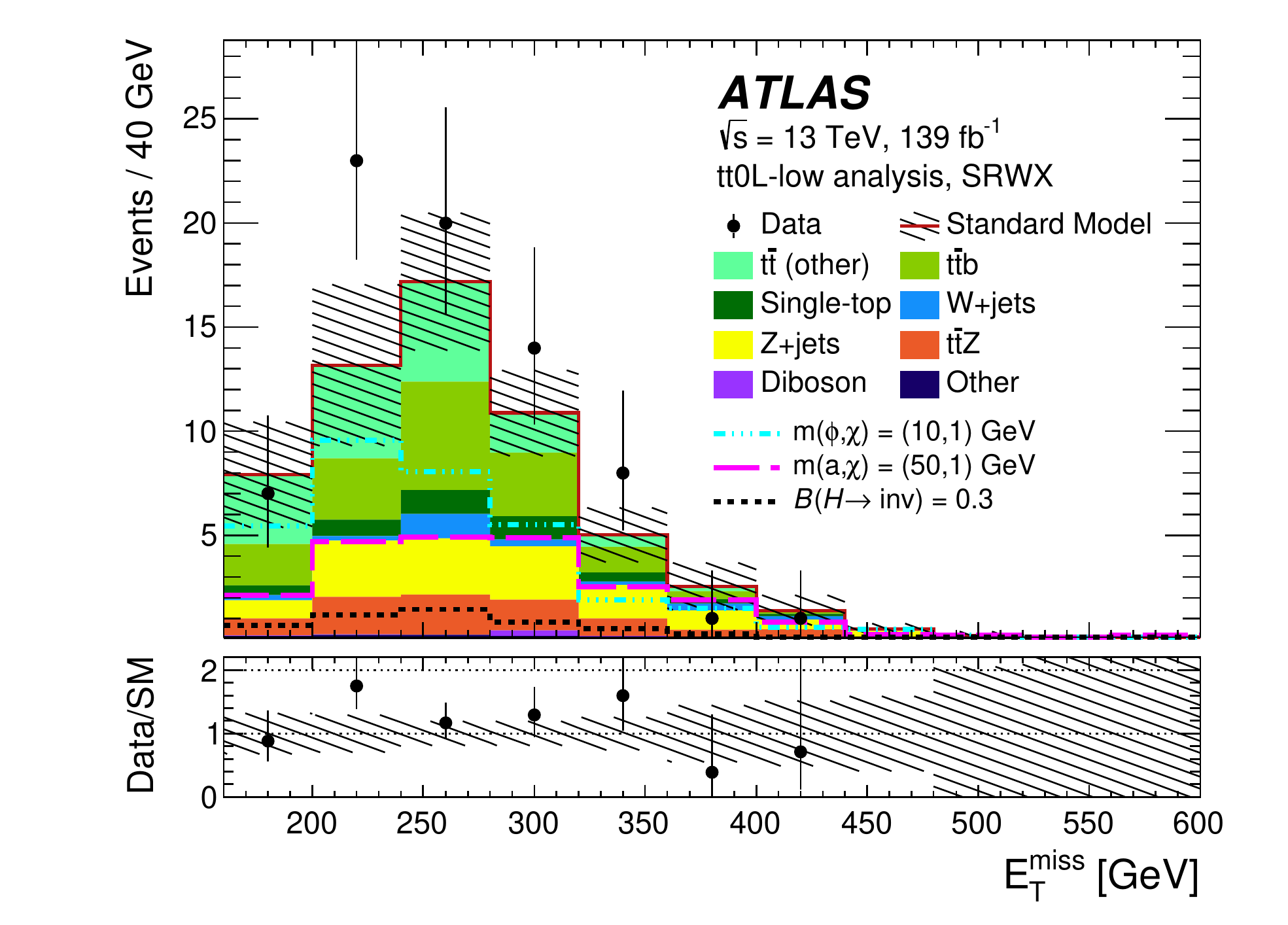}
\caption{}
\end{subfigure}
 
\begin{subfigure}[t]{0.5\textwidth}
\includegraphics[width=0.99\textwidth]{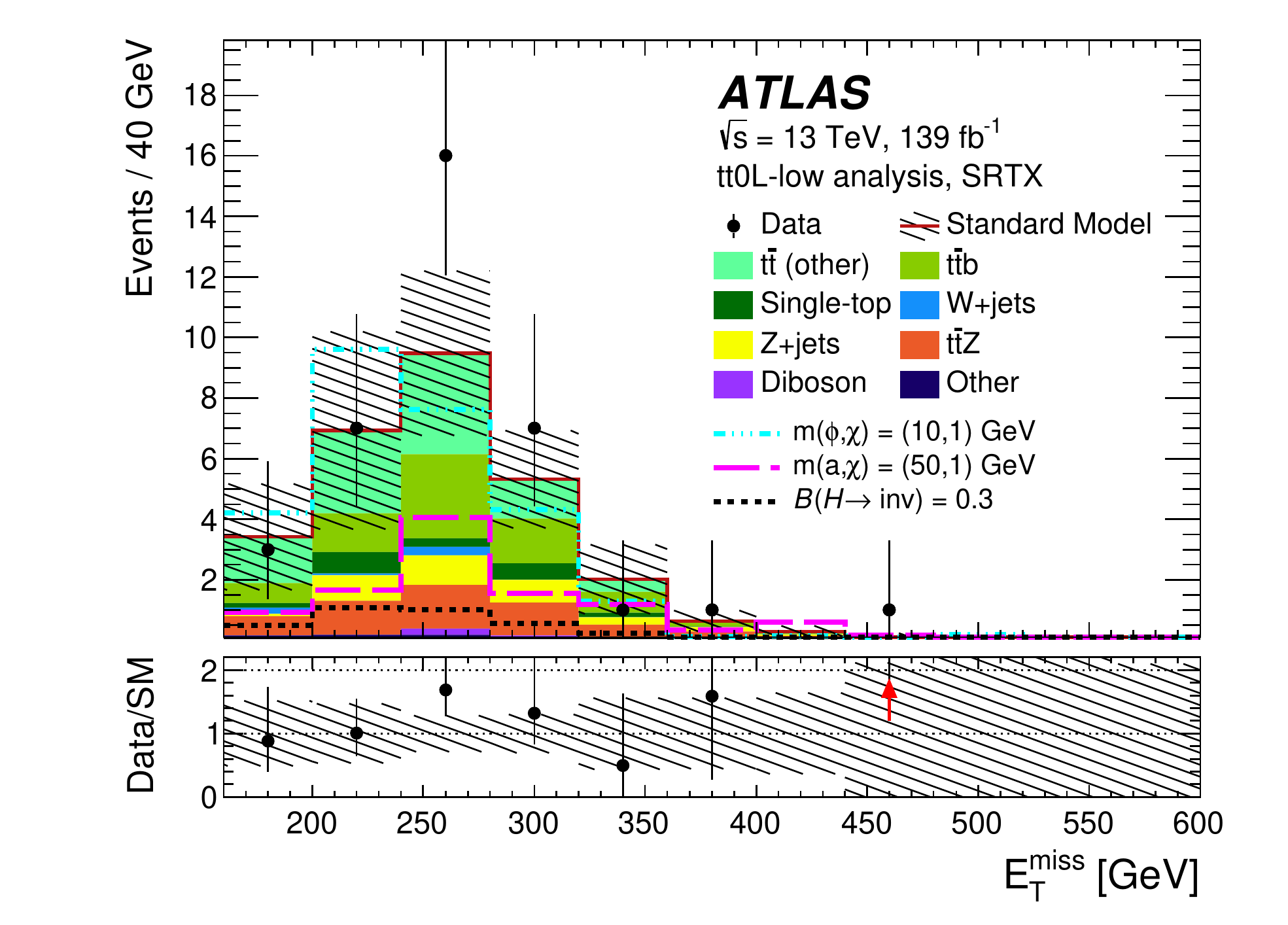}
\caption{}
\end{subfigure}
\caption{\met distributions in (a) SR0X, (b) SRWX and (c) SRTX events passing all the SR requirements.
The contributions from all SM backgrounds are shown after the profile likelihood simultaneous fit to all \noLlow\ CRs,
with the hatched bands representing the total uncertainty.
The category `\ttbar\ (other)' represents \ttbar\ events without extra jets or events with extra light-flavour jets.
`Other' includes contributions from $\ttbar{+}W$, $tZ$ and $tWZ$ processes.
The expected distributions for selected signal models are shown as dashed lines.
The overflow events are included in the last bin.
The bottom panels show the ratio of the observed data to the total SM background prediction,
with the hatched area representing the total uncertainty in the background prediction
and the red arrows marking data outside the vertical-axis range.}
\label{fig:lowmetpostfit}
\end{center}
\end{figure}

Figure~\ref{fig:lowmetpostfit} shows the \met distributions in the three \noLlow\ signal regions. The background contributions are obtained
from the profile likelihood simultaneous fit to all \noLlow\ CRs with a background-only hypothesis.
 
\begin{figure}[ht!b]
\begin{center}
\begin{subfigure}[t]{0.48\textwidth}
\centering
\includegraphics[width=0.99\textwidth]{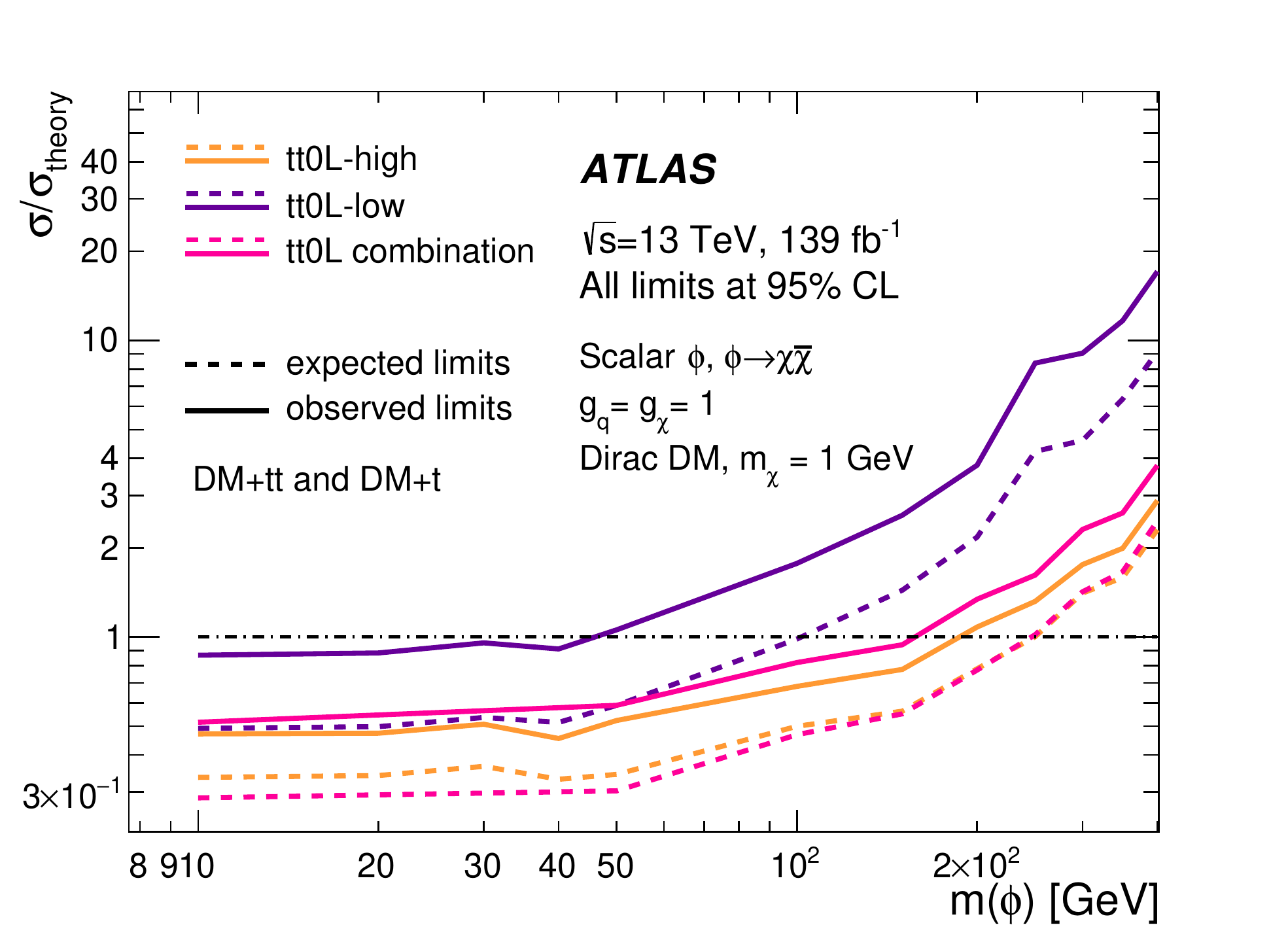}
\caption{}
\label{res0Lscalar}
\end{subfigure}
\begin{subfigure}[t]{0.48\textwidth}
\centering
\includegraphics[width=0.99\textwidth]{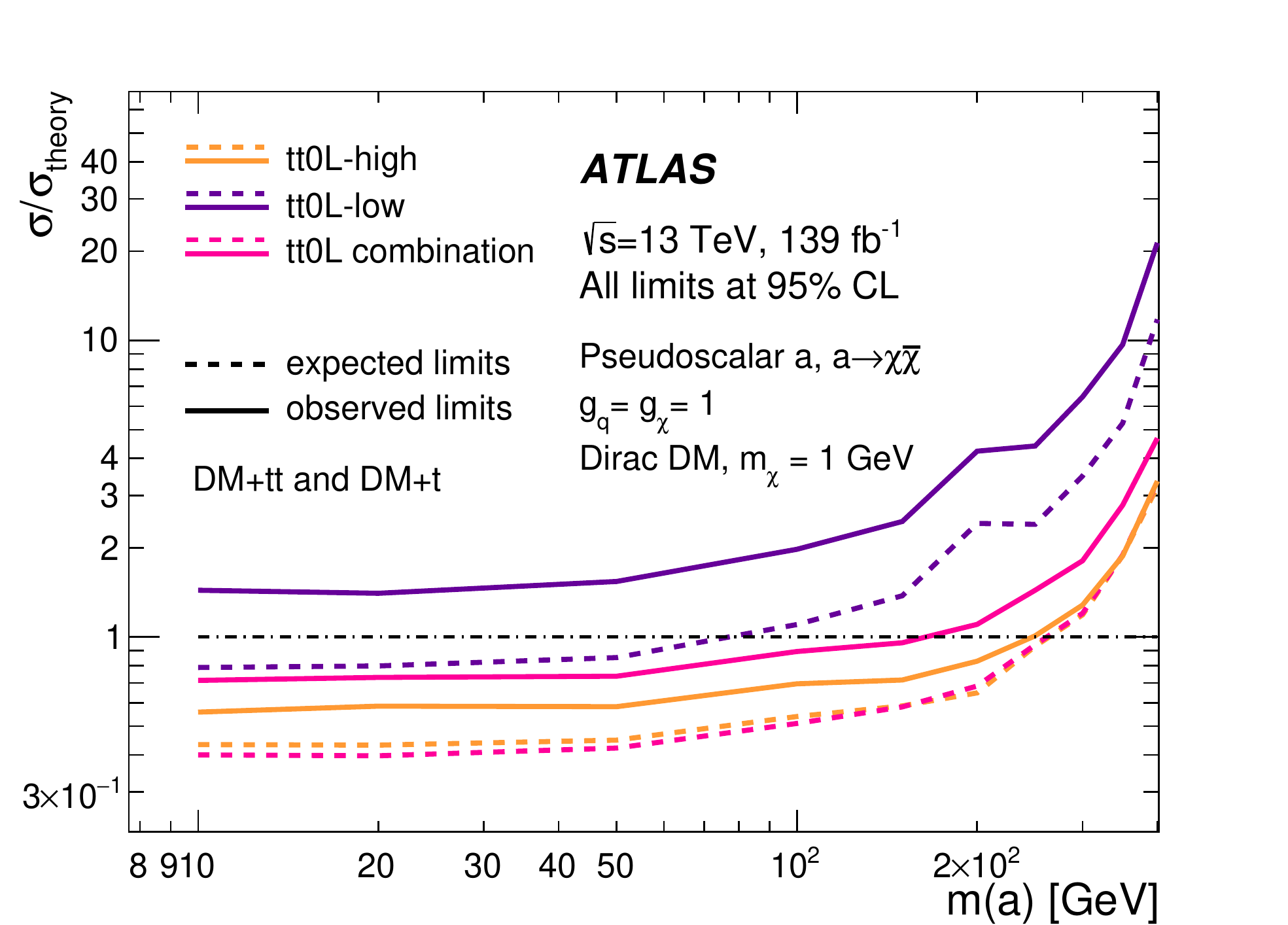}
\caption{}
\label{res0Lpseudo}
\end{subfigure}
\caption{Exclusion limits for colour-neutral (a) scalar or (b) pseudoscalar mediator
dark matter models as a function of the mediator mass
$m(\phi)$ or $m(a)$ for a DM mass $m_{\chi} =  1~\GeV$. Associated
production of DM with both single top quarks ($tW$ and $tj$
channels) and top quark pairs is considered. The limits are
calculated at 95\% CL and are expressed in terms of the ratio
of the excluded cross section to the nominal cross section for
a coupling assumption of $g = g_q = g_{\chi} = 1$.  The solid
(dashed) lines show the observed (expected) exclusion limits
for the \noLhigh\ and \noLlow\ analyses and their statistical combination.}
\end{center}
\end{figure}

Exclusion limits at 95\% CL are presented in Figures~\ref{res0Lscalar} and \ref{res0Lpseudo} for DM models with a spin-0 scalar or pseudoscalar mediator particle,
respectively. The \noLlow\ analysis, the \noLhigh\ analysis and the full \noL\ combination are presented separately in order to quantify the
improvement gained by adding the \noLlow\ channel to the \noL\ search. As they were designed to do, the \noLlow\ signal regions extend the sensitivity to
low-mass mediator models, with an improvement of up to about $15\%$ in the cross-section limit for scalar mediator particles.
 
\begin{figure}[ht!b]
\begin{center}
\begin{subfigure}[t]{0.48\textwidth}
\centering
\includegraphics[width=0.99\textwidth]{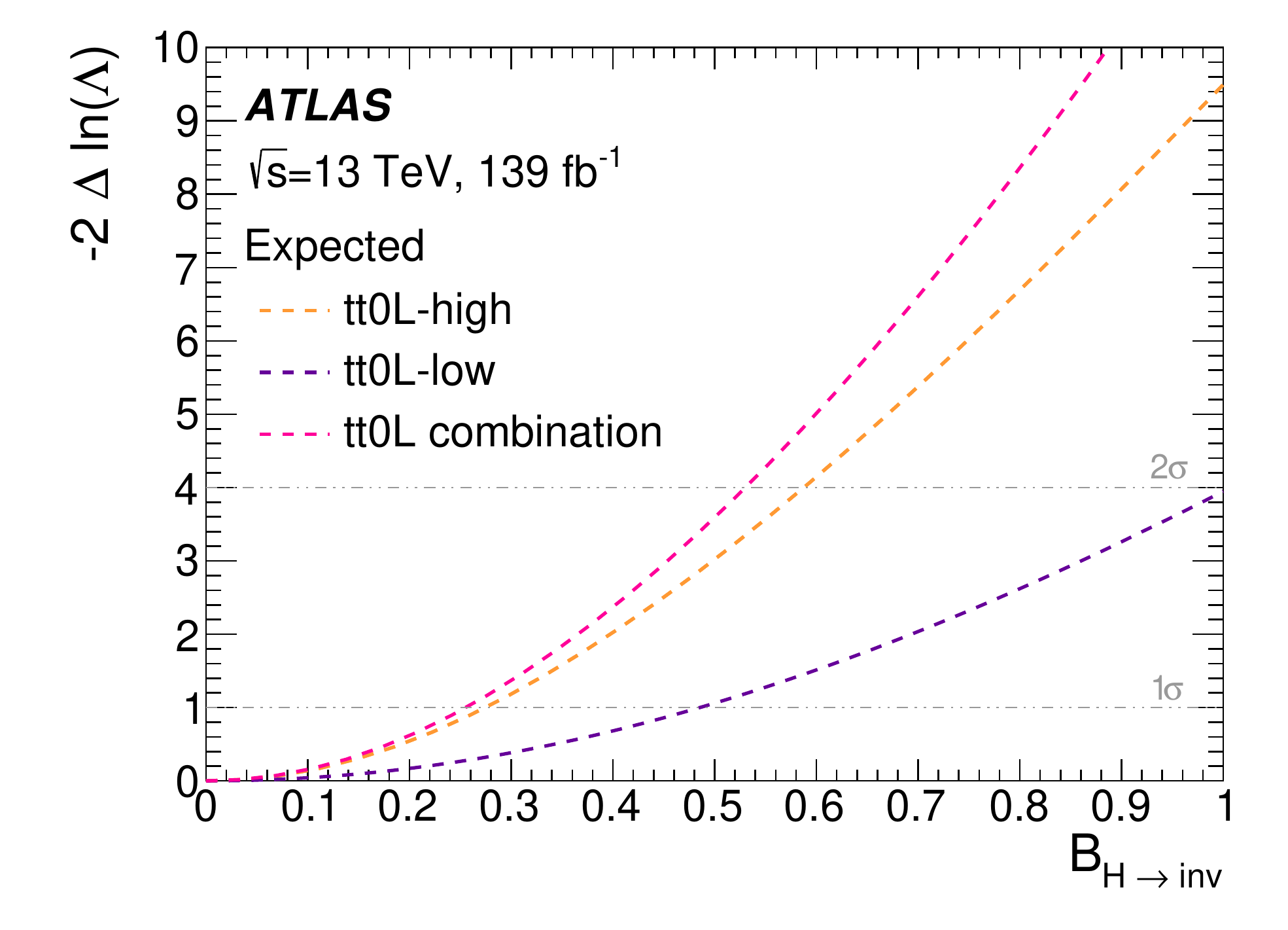}  
\caption{}
\end{subfigure}
\begin{subfigure}[t]{0.48\textwidth}
\centering
\includegraphics[width=0.99\textwidth]{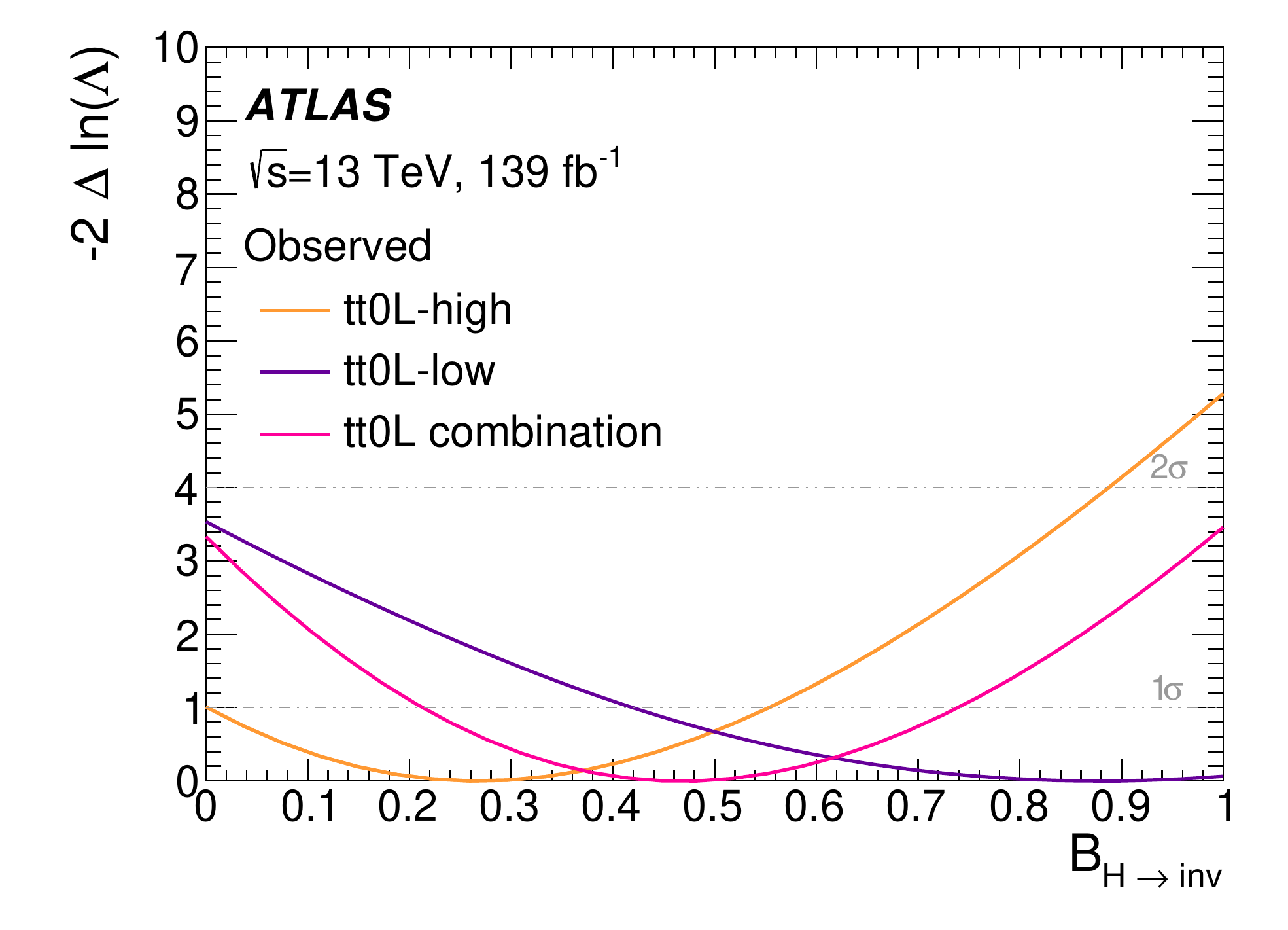}  
\caption{}
\end{subfigure}
\end{center}
\caption{(a) The expected negative logarithmic profile likelihood
ratios $-2\,\Delta\!\ln(\Lambda)$ as a function of \brinv for
each of the two tt0L analyses and their statistical combination
and (b) these likelihood ratios for the observed data.
\label{fig:lhscan_0L}
}
\end{figure}

In addition, the negative logarithmic profile likelihood ratios
$-2\,\Delta\!\ln(\Lambda)(\brinv;\theta)$ as a function of \brinv for the \noLlow\ and \noLhigh\ analyses,
and their combination, are illustrated in Figure~\ref{fig:lhscan_0L}.
 
Table~\ref{tab:DMtt_results_0L} presents the best-fit value, and the observed and expected upper limits on \brinv at the 95\% CL for
the \noLlow\ analysis, the \noLhigh\ analysis and their statistical combination. Since the \noLlow\ selection was designed to target mediator masses below 100~\GeV,
the improvement in the expected upper limit at the Higgs boson mass is found to be relatively small.
 
\begin{table} [htbp!]
\caption{
\label{tab:DMtt_results_0L}
Results from the \noLlow\ and \noLhigh\ searches for invisible decays of the
125~\GeV\ Higgs boson in the \tth topology using 139~\ifb\ of Run~2
data, and their statistical combination. Shown are the best-fit values
of \brinv, as well as observed and expected upper limits on \brinv at
the 95\% CL. The corresponding Asimov datasets for the expected
results are constructed using nuisance parameter values from a fit to
data with $\brinv = 0$, and the quoted uncertainty corresponds to the
68\% confidence interval.
}
\begin{center}
\begin{tabular}{l | c  c  c |l}
\hline
\multirow{2}{*}[1.5mm]{Analysis} & Best fit & Observed & Expected  &  \multirow{2}{*}[1.5mm]{Reference} \vspace{-3mm}\\
& \brinv & upper limit & upper limit & \\
\hline
\noLlow    &      $0.88^{+0.48}_{-0.46}$      &     1.80   & $1.09^{+0.50}_{-0.26}$  & this document                      \\
\noLhigh   &      $0.27^{+0.28}_{-0.27}$      &     0.80   & $0.59^{+0.29}_{-0.18}$  & \cite{SUSY-2018-12}, this document \\
\hline
tt0L comb. &      $0.48^{+0.27}_{-0.27}$      &     0.95   & $0.52^{+0.23}_{-0.16}$  & \cite{SUSY-2018-12}, this document \\
\hline
\end{tabular}
\end{center}
\end{table}

\FloatBarrier
 
\printbibliography

\clearpage
\input{atlas_authlist}

\end{document}

%% file: atlas_authlist.tex
 
\begin{flushleft}
\hypersetup{urlcolor=black}
{\Large The ATLAS Collaboration}

\bigskip

\AtlasOrcid[0000-0002-6665-4934]{G.~Aad}$^\textrm{\scriptsize 101}$,
\AtlasOrcid[0000-0002-5888-2734]{B.~Abbott}$^\textrm{\scriptsize 119}$,
\AtlasOrcid[0000-0002-7248-3203]{D.C.~Abbott}$^\textrm{\scriptsize 102}$,
\AtlasOrcid[0000-0002-1002-1652]{K.~Abeling}$^\textrm{\scriptsize 55}$,
\AtlasOrcid[0000-0002-8496-9294]{S.H.~Abidi}$^\textrm{\scriptsize 29}$,
\AtlasOrcid[0000-0002-9987-2292]{A.~Aboulhorma}$^\textrm{\scriptsize 35e}$,
\AtlasOrcid[0000-0001-5329-6640]{H.~Abramowicz}$^\textrm{\scriptsize 150}$,
\AtlasOrcid[0000-0002-1599-2896]{H.~Abreu}$^\textrm{\scriptsize 149}$,
\AtlasOrcid[0000-0003-0403-3697]{Y.~Abulaiti}$^\textrm{\scriptsize 116}$,
\AtlasOrcid[0000-0003-0762-7204]{A.C.~Abusleme~Hoffman}$^\textrm{\scriptsize 136a}$
% ...
\bigskip
\\

% ...
$^{29}$Physics Department, Brookhaven National Laboratory, Upton NY; United States of America.\\
% ...
$^{35}$$^{(a)}$Facult\'e des Sciences Ain Chock, R\'eseau Universitaire de Physique des Hautes Energies - Universit\'e Hassan II, Casablanca;$^{(b)}$Facult\'{e} des Sciences, Universit\'{e} Ibn-Tofail, K\'{e}nitra;$^{(c)}$Facult\'e des Sciences Semlalia, Universit\'e Cadi Ayyad, LPHEA-Marrakech;$^{(d)}$LPMR, Facult\'e des Sciences, Universit\'e Mohamed Premier, Oujda;$^{(e)}$Facult\'e des sciences, Universit\'e Mohammed V, Rabat;$^{(f)}$Institute of Applied Physics, Mohammed VI Polytechnic University, Ben Guerir; Morocco.\\
% ...
$^{55}$II. Physikalisches Institut, Georg-August-Universit\"{a}t G\"ottingen, G\"ottingen; Germany.\\
% ...
$^{101}$CPPM, Aix-Marseille Universit\'e, CNRS/IN2P3, Marseille; France.\\
$^{102}$Department of Physics, University of Massachusetts, Amherst MA; United States of America.\\
% ...
$^{116}$Department of Physics, New York University, New York NY; United States of America.\\
% ...
$^{119}$Homer L. Dodge Department of Physics and Astronomy, University of Oklahoma, Norman OK; United States of America.\\
% ...
$^{136}$$^{(a)}$Departamento de F\'isica, Pontificia Universidad Cat\'olica de Chile, Santiago;$^{(b)}$Millennium Institute for Subatomic physics at high energy frontier (SAPHIR), Santiago;$^{(c)}$Instituto de Investigaci\'on Multidisciplinario en Ciencia y Tecnolog\'ia, y Departamento de F\'isica, Universidad de La Serena;$^{(d)}$Universidad Andres Bello, Department of Physics, Santiago;$^{(e)}$Instituto de Alta Investigaci\'on, Universidad de Tarapac\'a, Arica;$^{(f)}$Departamento de F\'isica, Universidad T\'ecnica Federico Santa Mar\'ia, Valpara\'iso; Chile.\\
% ...
$^{149}$Department of Physics, Technion, Israel Institute of Technology, Haifa; Israel.\\
$^{150}$Raymond and Beverly Sackler School of Physics and Astronomy, Tel Aviv University, Tel Aviv; Israel.\\
% ...
$^{a}$ Also Affiliated with an institute covered by a cooperation agreement with CERN.\\
$^{b}$ Also at Borough of Manhattan Community College, City University of New York, New York NY; United States of America.\\
$^{c}$ Also at Bruno Kessler Foundation, Trento; Italy.\\
$^{d}$ Also at Center for High Energy Physics, Peking University; China.\\
$^{e}$ Also at Centro Studi e Ricerche Enrico Fermi; Italy.\\
$^{f}$ Also at CERN, Geneva; Switzerland.\\
% ...

\end{flushleft}
